\documentclass[trackchanges]{aastex701}

\begin{document}

\title{The SOL \textit{(Solar Origin and Life)} Project: \\ Detailed characterization of candidates for the ZAMS and Subgiant stages}

\author[orcid=0009-0001-4152-1342,gname=Eduardo-Oliveira, sname='C.S.']{Eduardo-Oliveira C.S.}
\affiliation{Observatório do Valongo, Universidade Federal do Rio de Janeiro, Ladeira do Pedro Antonio, 43, 20080-090 Rio de Janeiro, Brazil}
%\altaffiliation{Universidade Federal do Rio de Janeiro (UFRJ)}
\email[show]{carlos17@ov.ufrj.br}  

\author[orcid=0000-0002-9089-0136, gname=Ghezzi, sname='Luan']{Ghezzi, L.} 
\affiliation{Observatório do Valongo, Universidade Federal do Rio de Janeiro, Ladeira do Pedro Antonio, 43, 20080-090 Rio de Janeiro, Brazil}
\affiliation{Observatório Nacional, MCTIC (ON), Rua Gal. José Cristino 77, São Cristóvão, 20921-400, Rio de Janeiro, Brazil}
%\altaffiliation{Universidade Federal do Rio de Janeiro (UFRJ)}
\email{luanghezzi@ov.ufrj.br}

\author[orcid=0000-0003-0911-1324, gname=Porto de Mello, sname='Gustavo Frederico']{Porto de Mello, G. F.} 
\affiliation{Observatório do Valongo, Universidade Federal do Rio de Janeiro, Ladeira do Pedro Antonio, 43, 20080-090 Rio de Janeiro, Brazil}
%\altaffiliation{Universidade Federal do Rio de Janeiro (UFRJ)}
\email{gustavo@ov.ufrj.br}

\author[orcid=0000-0002-1387-2954, gname=Lorenzo-Oliveira, sname='Diego']{Lorenzo-Oliveira, D.} 
\affiliation{Laboratório Nacional de Astrofísica, Rua Estados Unidos 154, 37504-364, Itajubá - MG, Brazil}
%\altaffiliation{Ministério da Ciência, Tecnologia e Inovação (MCTI)}
\email{diegolorenzo.astro@gmail.com}

\author[orcid=0009-0009-3820-1346, gname=Souza dos Santos, sname='Paulo Vitor']{Souza dos Santos, P. V.} 
\affiliation{Observatório do Valongo, Universidade Federal do Rio de Janeiro, Ladeira do Pedro Antonio, 43, 20080-090 Rio de Janeiro, Brazil}
%\altaffiliation{Universidade Federal do Rio de Janeiro (UFRJ)}
\email{paulovss@ov.ufrj.br}

\author[orcid=0000-0002-1549-626X, gname=Costa-Almeida, sname='Ellen']{Costa-Almeida, E.} 
\affiliation{Observatório Nacional, MCTIC (ON), Rua Gal. José Cristino 77, São Cristóvão, 20921-400, Rio de Janeiro, Brazil}
%\altaffiliation{Ministério da Ciência, Tecnologia e Inovação (MCTI)}
\email{ellenalmeida@on.br}

\begin{abstract}

The context of the Sun in the galactic neighborhood is not well understood, especially when we compare its physical properties to those of nearby stars. Thereby, we still cannot fully comprehend whether or not the Sun is a typical star. This work aims to identify and characterize stars aligned with the solar evolutionary track that could represent it at the ZAMS and subgiant stages. We performed a spectroscopic analysis of 18 photometrically selected candidates using high-resolution and high-SNR spectra as well as the classical spectroscopic method, based on the excitation and ionization equilibria of Fe I and Fe II lines. Additionally, we derived evolutionary parameters using isochrones, and kinematic parameters. We also estimated chromospheric activity levels and performed age estimates through 3 additional independent methods: activity-age relations using the \ion{Ca}{2} H \& K and H$\alpha$ lines, and rotation periods estimated from TESS light curves. We identified three candidates that provide a good match to the Sun at $\approx$ 0.5 Gyr (HD 13531 and HD 61033) and subgiant (HD 148577) stages. Moreover, HD 197210 could be of interest when studying the Sun at $\approx$2 Gyr, when the Earth’s atmosphere started having a significant amount of oxygen. Our selection method was successful and we were able to identify stars similar to the Sun at different evolutionary stages, which is essential for future research in the search of exoplanets and understand habitability, especially with the advent of the next generation of exoplanet-hunting instruments. 

\end{abstract}



\section{Introduction}
\label{introduction}

The Sun is currently the only star known to harbor a planet onto which life flourished. This striking feature could be a direct result of many possible peculiarities it exhibits when compared to stars in the solar neighborhood. The Sun's location possibly minimizes the number of passages through the spiral arms, decreasing the probabilities of exposure to close-by supernova events or encounters with giant molecular clouds that could disturb the Solar System. Therefore, the Sun's location in the Galaxy might have been fundamental to the long-term maintenance of a suitable biosphere for life on Earth as well as to limit the amount of mass extinction events \citep{leitch1998mass,2006AsBio...6..308P,2009ASPC..420..349P}. However, its birthplace could be located at a Galactocentric radius $R_{g} \approx$ 5 - 6 kpc followed by an outward migration to its current position \citep{nieva2012present} close to galactic corotation \citep{lepine2001new}.

Furthermore, there was early evidence that the Sun is metal-rich for its age \citep{10.1093/mnras/279.2.447}. However, \citet{casagrande2011new} presented a new revision of the Geneva-Copenhagen survey and their results suggest that the Sun is indeed a typical star when we compare its metallicity to the distribution of metallicities of the solar neighborhood. Other possible chemical peculiarities have been suggested for the Sun, such as a higher C/O ratio relative to the cosmic abundance standard (CAS) (\cite{nieva2012present}, but see also \citealt{asplund2009chemical}), a larger Li depletion relative to solar-type stars (e.g., \citealt{pasquini1994lithium,2007A&A...468..663T,ramirez2012lithium,2025A&A...700A.127C}; but see also \citealt{Ghezzi_2010b}) and deficiency in refractory elements relative to volatiles when compared to the so-called solar twin stars (e.g., \citealt{Mel_ndez_2009}; \citealt{10.1093/mnras/staa578}; \citealt{2024ApJ...965..176R}). The first feature influences the structure of the planets that will be formed in the system, while the second and the third could be a consequence of giant or terrestrial planet formation, respectively. However, the effect of depletion of refractory elements could also be reproduced through a combination of multiple factors that may have acted simultaneously (such as the chemical evolution of the Galaxy, the Sun’s possible migration, and processes intrinsic to the protoplanetary disk), making it very difficult to independently quantify their individual contributions \citep{2025A&ARv..33....3G}.

There is also some evidence that the Sun might have a longer rotational period (\citealt{pace2004age}; but see also \citealt{dos2016solar}) and a lower chromospheric activity level \citep{hall2000evidence} than stars of similar age, both of which are possibly connected with the enhanced Li depletion. However, recent studies suggest that the Sun is not an uncommon star, considering its age and spectral type \citep{hall2007sun, doi:10.1126/science.aay3821}. Nonetheless, its activity cycles seem smoother and more regular when compared to Sun-like stars \citep{radick2018patterns}. This factor could have played a significant role in the habitability of our Solar System.

More recently, \citet{10.1093/mnras/staf1149} performed a comprehensive statistical analysis to search for anomalous properties that the Sun may exhibit within a large sample of nearby stars (up to 25 pc), including solar analogs and twins. He identified its mass (top $\approx$8 per cent), low photometric variability on short timescales (bottom $\approx$0.2 per cent), specific light and heavy element abundance patterns, slow rotation, and low superflare rate as its most peculiar characteristics.

It is thus clear that an investigation of the solar properties relative to similar stars and a detailed description of its evolution are extremely important for understanding how life appeared and evolved on Earth as well as for the searches for habitable planets and extraterrestrial life. For this reason, our group has been analyzing stars that follow the solar evolutionary track on the HR diagram to search for proxies that could represent the Sun at different stages of its evolution. The first results of this effort were the identification of the solar twin 18 Sco (HD 146233, HR 6060; \citealt{de1997hr}), which persists as one of the best candidates found to date. More recently, we identified 10 new likely solar twins that are being subjected to more detailed analyzes \citep{de2014photometric}. These stars add up to a much larger sample of candidates to solar analogs and twins that were proposed by different groups over the past decades \citep[e.g.,][]{hardorp1978sun,cayrel1981search,cayrel1996stars,soubiran2004top,king2005keck,2007A&A...468..663T,Mel_ndez_2009,onehag2011m67,ramirez2014solar, 10.1093/mnras/stab987,lehmann2022survey}.

Since stars that resemble the current Sun do not provide the entire picture about its evolution, the searches described above were extended to other evolutionary stages along a solar evolutionary track. \citet{ribas2010evolution} presented a detailed analysis of $\kappa^{1}$ Cet, which is an analog of the Sun when life appeared on Earth. \citet{do2013future} analyzed CoRoT ID 102684698, which is an evolved solar twin that could provide insights into the future of the Sun. Solar proxies in different evolutionary stages were also found by the Sun in Time program \citep[e.g.,][]{dorren1994hd,gudel1997x,ribas2005evolution,ribas2010evolution} as well as by other groups \citep{dravins1993a,dravins1993b,dravins1993c,gaidos1998nearby, Galarza_2025}.

In order to continue and extend previous initiatives, we started the \textit{Solar Origin and Life} (SOL) project back in 2003. The main goal of this effort is to identify stars that resemble the Sun along its evolutionary path in order to understand how its properties influenced the formation and evolution of the Solar System and life on Earth. In this paper, we present the pilot study for this project and the resulting candidates we identified. Section \ref{sec:sample_selection_data} describes the sample selection, observations and data reduction. Section \ref{sec:stellar_parameters} contains the determination of the atmospheric, evolutionary and kinematical parameters, as well as the activity indicators in Section \ref{sec:stellar_activity}. In Section \ref{best_candidates}, we discuss the results and use them to carefully select the best candidates to represent the Sun. Our concluding remarks are presented in Section \ref{sec:conclusions}.

%% The "ht!" tells LaTeX to put the figure "here" first, at the "top" next
%% and to override the normal way of calculating a float position.
%% The asterisk after "figure" tells the compiler to span multiple columns
%% if a two column style is selected.
%\begin{figure*}[ht!]
%\plotone{AuthorChargeInfographic.png}
%\caption{The AAS journals are operated as a nonprofit venture, and author charges fairly recapture costs for the services provided in the publishing process. The chart above breaks down the services that author charges go toward. The AAS Journals' Business Model is outlined in a \href{https://aas.org/posts/news/2023/08/aas-open-access-publishing-model-open-transparent-and-fair}{2023 post}.
%\label{fig:general}}
%\end{figure*}

\section{Sample selection and data}
\label{sec:sample_selection_data}
\subsection{Sample Selection}
\label{sec:sample_selection} % used for referring to this section from elsewhere

Any study that attempts to select solar proxies which lie along or close to its evolutionary path will be naturally model dependent. We decided to adopt as our reference the evolutionary track from \cite{schaller1992new} that was calculated for 1 M$_{\odot}$ and Z = 0.02. These models do not take overshooting and rotation into account and had to be adjusted so that a point with an age of 4.56 Gyr (adopted from \citealt{2012Sci...338..651C}) corresponded to the current solar effective temperature ($T_{eff}$ $\approx$ 5780 K) and luminosity. Although more recent evolutionary tracks are available in the literature, we opted for this grid because its uses a similar physics as the one employed by \cite{sackmann1993} and we wanted to benefit from the wealth of information contained in this latter study. Moreover, we wanted to keep this pilot study consistent with previous selections performed by our group (e.g., \citealt{de2014photometric}). We note, however, that comparisons with similar but more recent evolutionary tracks \citep[e.g.,][]{demarque2004y2, chen2015parsec} did not reveal discrepancies that would significantly affect our results for the evolutionary stages analyzed here. We chose the evolutionary tracks of \cite{schaller1992new} to maintain consistency with \cite{de2014photometric}. We note that more recent evolutionary tracks are available such as PARSEC (\citealt{bressan2012parsec}) and Dartmouth (\citealt{2008ApJS..178...89D}); however, the differences in $T_{\mathrm{eff}}$ ($\Delta_{max} = 70 K $) are below the typical external uncertainties found in the comparisons of parameters for solar-type stars \citep[e.g.,][]{2016ApJS..226....4H} and the differences in luminosity ($\Delta_{max} = 0.03$) are consistent with our internal uncertainties (mean $\sigma_{L/L_{\odot}} = 0.02$ for the ZAMS and mean $\sigma_{L/L_{\odot}} = 0.05$ for the SG candidates), as will be shown in Sections \ref{sec:atmparam} and \ref{sec:evolparam}.

Following the detailed discussion performed in Section 3.2 of \cite{sackmann1993}, we chose two stages, Zero Age Main Sequence (ZAMS) and Subgiant (SG), along our reference solar evolutionary track that are related to the points in their Figure 2. Their corresponding effective temperatures and luminosities are given in Table \ref{tabela_param}, in which we also show the parameters of the Present Sun (as given by \citealt{schaller1992new}) for comparison. The stage in which the Sun enters on the main sequence is represented by the point ZAMS (Zero Age Main Sequence) and it was chosen according to point A in Figure 2 of \cite{sackmann1993}. We can see that it is cooler, less luminous and smaller than the current Sun, shown by the point Today. The stage in which the Sun leaves the main sequence is depicted by the point TO (turn-off). At this moment, the Sun will be slightly cooler, but $\approx$34\% larger and $\approx$76\% more luminous.

The selection of stars that could represent the Sun in these two evolutionary stages (ZAMS and SG) was performed by \citet{ghezzi2005} and relies on the photometric boxes in the M$_{V_T}$ - (B$_{T}$-V$_{T}$) plane, as described by \citet{de2014photometric}. Briefly, we converted the luminosities in Table \ref{tabela_param} to absolute visual magnitudes M$_{V_T}$ in the V$_{Tycho}$ band using M$_{bol_\odot}$ = 4.81 (as adopted in \citealt{de2014photometric}) and bolometric corrections from \citet{habets1981empirical}. The color indexes (B$_{T}$-V$_{T}$) were calculated using the photometric calibration from \citet{ghezzi2005}, which was determined following a similar procedure as the one described in the Appendix A of \citet{de2014photometric}. The inputs to the calibration are the effective temperatures, taken from Table \ref{tabela_param}, and the solar metallicity, which was set to [Fe/H] = 0.00 by definition. The values for M$_{V_T}$ and (B$_{T}$-V$_{T}$) for the Sun ZAMS and SG are shown in Table \ref{tabela_param}.

In order to set the widths of the photometric boxes around these points, we need their uncertainties. These were estimated in the same way as described by \citet{de2014photometric}. As we decided to perform the cut V$_{T}$ $<$ 8.1 in order to select only the brightest stars for our pilot study, the values of these uncertainties are $\sigma$(M$_{V_T}$) = 0.07 and $\sigma$(B$_{T}$-V$_{T}$) = 0.013 (\citealt{de2014photometric}). We constructed our photometric boxes considering 2$\sigma$ intervals around the photometric points for the Sun ZAMS and SG. Finally, we searched the Hipparcos catalog (\citealt{perryman1997hipparcos}) for stars within these photometric boxes that also had V$_{T}$ $<$ 8.1; since our selection was carried out before the launch of Gaia \citep{gaia2018gaia}, we decided to use Hipparcos; future selections of our project will make use of Gaia data.

We selected 20 candidates but removed two stars from our sample. One of them (HD 114260) is a spectroscopic binary\footnote{https://simbad.u-strasbg.fr/simbad/sim-basic?Ident=HD+114260} and the determination of spectroscopic parameters might be affected by the contamination of the spectral lines from the secondary. 
For the other one (HD 215028), we were not able to observe or retrieve from public archives a spectrum with the minimum quality required (see Section \ref{sec:data}) for our detailed analysis. Our final sample contains 18 stars, 8 of which are candidates to the ZAMS stage and 10 to the SG stage. Furthermore, we decided to include the solar twin 18 Sco (HD 146233) as a control star in the sample, as it is well studied in the literature (e. g.,\citealt{de1997hr}, \citealt{bazot2011radius}, \citealt{ramirez2014solar}, \citealt{bazot2018modelling}, \citealt{do2023hale}), and we can use it to check the precision of our method.

\subsection{Data and reduction}
\label{sec:data}

The data for our sample stars were obtained from two main sources. The spectra of 2 stars were obtained using the MUSICOS \textit{echelle} spectrograph attached to the 1.6m telescope at Pico dos Dias Observatory (OPD) between 2013 and 2017. The Pico dos Dias Observatory and MUSICOS are administered by Laboratório Nacional de Astrofísica (LNA) and its resolving power is aproximatelly R $\approx$ 35000 with a spectral coverage of 5400-8800\AA. We reduced the MUSICOS spectra using standard procedures (bias, flat-field and scattered light corrections; one dimensional extraction and wavelength calibration).

The other main source was public databases from the following spectrographs: FEROS/ESO (\citealt{kaufer1997feros}), UVES/ESO (\citealt{2000SPIE.4008..534D}), HARPS-S/ESO (\citealt{2003Msngr.114...20M}), HARPS-N/TNG (\citealt{10.1117/12.925738}), HIRES/KECK (\citealt{vogt1994hires}) and ESPaDOnS/CFHT (\citealt{donati2003espadons}). In addition to the MUSICOS's spectra, we were able to retrieve high-quality spectra for our candidates with resolving powers ranging from 35000 up to 115000 and minimum signal-to-noise ratio (SNR) of 100. We acquired almost 90 high-quality spectra for our 18 candidates and 18 Sco, which resulted in a better estimation of the fundamental parameters of the candidates. The public spectra were already reduced by their own automated pipelines.

%\textbf{We do not expect the different resolutions and SNR intervals of our spectra to have affected our analysis for two main reasons: the line list used in the analysis was carefully selected, prioritizing lines located in well-behaved continuum regions and avoiding blended lines; and the Sol Project sample is highly homogeneous due to the selection method employed. This can be observed in Table \ref{table_param_atm} where our estimated internal errors are very precise.}

We corrected all spectra for their Doppler shifts using the \textit{fxcor} and \textit{dopcor} tasks from IRAF (Image Reduction and Analysis Facility)\footnote{https://iraf-community.github.io/} and a solar rest-frame spectrum as reference (\citealt{2000vnia.book.....H}). Finally, we estimated their SNR values using IRAF's \textit{bplot} routine on 143 continuum regions carefully selected using the solar flux spectrum from \citet{2000vnia.book.....H}. After perfoming 2 rounds of 2$\sigma$ clipping we established their mean value as the final one for each spectra (see Table \ref{table_param_input}). We note that although our spectroscopic data are not homogeneous, this did not affect the internal consistency of our results (see Section \ref{sec:atmparam}).

\begin{table}[h!]
    \centering
    \caption{Values adopted from \citet{sackmann1993} for the evolutionary stages discussed in this work. The canonical values of the present Sun were included for comparison purposes.}
    \label{tabela_param}
    \begin{tabular}{lcccccc}
        \hline
        Stage & $T_{eff}$ & L & R & Age &  (M$_{V_{T}}$) &  (B$_{T}$-V$_{T}$)\\
        &  (K) & (L$_{\odot}$) & (R$_{\odot}$) & (Gyr) & & \\
        \hline
        ZAMS & 5586 & 0.70 & 0.897 & 0.0 & 5.36 & 0.81 \\
        Present Sun & 5777 & 1.00 & 1.000 & 4.5 & 4.88 & 0.73\\
        SG & 5743 & 1.76 & 1.343 & 10.0 & 4.30 & 0.75\\
        \hline
    \end{tabular}
\end{table}

\begin{longrotatetable}
\begin{table}
    \centering
    \caption{All input parameters of the sample stars that were used throughout the analysis. The columns in the table represent: ID, spectral type, the evolutionary stage for which the candidate was selected, parallax, magnitude in the V band, RA and DEC coordinates, proper motion in RA and DEC, radial velocity, number of spectra analyzed, source instruments and the SNR interval.}
    \begin{tabular}{cccccccccccccc}
        \hline
        Star          & Spectral type & Candidate & $\pi$ & V & RA & Dec & $\mu_{\alpha}$ & $\mu_{\delta}$ & V$_{rad}$ & N$_{spectra}$ & Instrument(s)&  SNR interval\\
        & & & ($mas$) & (mag) & (degrees) & (degrees) & ($mas/yr$) & ($mas/yr$) & $km s^{-1}$ & & &\\
        \hline
        HD 13531 & G7V & ZAMS & 38.27 $\pm$ 0.06 & 7.36 $\pm$ 0.01 & 33.31 & 40.51 & 59.262 & -90.533 & 6.70 & 3 & ESPaDOnS & 286 - 321\\
        HD 21411 & G8V & ZAMS & 34.30 $\pm$ 0.04 & 7.88 $\pm$	0.01 & 51.55 &-30.62 & 218.073 & 226.095 &  17.80 & 3 & HARPS-S & 113 - 127\\
        HD 25918 & G5  & ZAMS & 30.76 $\pm$ 0.06 & 7.72 $\pm$	0.01 & 62.06 & 44.66 & -298.823 & -133.236  & -36.15 & 3 & SOPHIE & 165 - 168\\
        HD 55720 & G8V & ZAMS & 36.21 $\pm$ 0.03 & 7.47 $\pm$	0.01 & 107.88 & -49.42 & -7.129 & 817.422 & 87.41 & 1 & FEROS & 397\\
        HD 61033 & G7V & ZAMS & 36.06 $\pm$ 0.96  & 7.57 $\pm$	0.01 & 113.62 & -52.97 &-25.532  & 271.951 & 14.98 & 1 & FEROS & 242\\
        HD 64114 & G7V & ZAMS & 31.69 $\pm$ 0.04 & 7.72 $\pm$	0.01 & 117.98 &	-11.03 & -69.453  & -173.481 & 53.18 & 1 & FEROS & 370\\
        HD 182619 & G5 & ZAMS & 30.53 $\pm$ 0.04 & 7.80 $\pm$ 0.01 & 291.17 & 22.20  & 134.856 & -83.177 & 8.27 & 3 & SOPHIE & 149 - 152\\
        HD 197210 &G5V & ZAMS & 33.50 $\pm$ 0.04 & 7.61 $\pm$ 0.01 & 310.62 & -5.30  & -60.414  & -172.389 & 3.22 & 6 & HARPS-N, HARPS-S & 133 - 138, 170 - 191\\
        \hline
        HD 15942 & G0  & SG & 22.39 $\pm$ 0.04 & 7.50 $\pm$	0.01 & 38.52 & 12.18  & 183.713 & -35.560 & 34.24 & 3 & HARPS-S & 183 - 186\\
        HD 19308 & G0  & SG & 25.70 $\pm$ 0.07 & 7.37 $\pm$	0.01 & 46.91 & 36.62  & 241.022 & -217.657 & 32.74 & 3 & SOPHIE & 174 - 208\\
        HD 24040 & G1  & SG & 21.42 $\pm$ 0.06 & 7.50 $\pm$	0.01 & 57.60 & 17.48  & 113.215 & -251.033 & -9.34 & 6 & SOPHIE, UVES & 149 - 154, 191 - 204\\
        HD 69809 & G0  & SG & 18.75 $\pm$ 0.05 & 7.86 $\pm$	0.01 & 124.81 &	14.20 & 35.895 & 24.442 & 17.47 & 5 & HIRES, UVES & 294 - 299, 192 - 248\\
        HD 74698 & G5V & SG & 19.18 $\pm$ 0.03 & 7.76 $\pm$	0.01 & 130.07 &	-71.88 & 106.122  & 61.270 & 38.75 & 3 & HARPS-S & 103 - 128\\
        HD 111398 & G5V& SG & 27.50 $\pm$ 0.05 & 7.10 $\pm$ 0.01 &	192.22 & 12.10 & 233.049  & -139.621 & 3.18 & 3 & SOPHIE & 166 - 178\\
        HD 148577 & G5V& SG & 18.52 $\pm$ 0.05 & 7.96 $\pm$ 0.01 & 247.43 & -18.68 & 81.435  & -238.526 & 62.74 & 4 & HARPS-S, MUSICOS & 65 - 66*, 117 - 173\\
        HD 175425 & G0 & SG & 18.56 $\pm$ 0.03 & 7.89 $\pm$ 0.01 & 283.41 & 37.99 & -46.687 & -84.223 & -67.74 & 1 & HIRES & 167\\
        HD 196050 & G3V& SG & 19.71 $\pm$ 0.04 & 7.49 $\pm$ 0.01 & 309.47 & -60.63 & -191.122 & -64.922 & 61.37 & 7 & FEROS, HARPS-S, & 298, 192 - 199,\\
         &  &  &  &  &  &  &  &  &  & & UVES &  184 - 282\\
        HD 213575 & G5V& SG & 26.73 $\pm$ 0.09 & 6.95 $\pm$ 0.01 & 338.14 & -6.47 & 291.969 & 30.281 & -21.50 & 12 & HARPS-N, HARPS-S, & 145 - 151, 207 - 214,\\
        &  &  &  &  &  &  &  &  &  & & SOPHIE, UVES &  168 - 191, 173 - 183\\
        \hline
        18 Sco   & G2V & TWIN & 70.77   $\pm$ 0.11 & 5.50 $\pm$ 0.01 & 243.91 & -8.37 & 232.159 & -495.368 & 11.76 & 20 & ESPaDOnS, FEROS, HARPS-N, & 215 - 237, 372 - 382, 280 - 286,\\
         &  &  &  &  &  &  &  &  &  & & HARPS-S, HIRES, SOPHIE, UVES, & 340 - 361, 351 - 487, 132 - 134, 190 - 281,\\
         &  &  &  &  &  &  &  &  &  & & MUSICOS & 266 - 191\\
        \hline
    \end{tabular}
\label{table_param_input}
\end{table}
\end{longrotatetable} %Tabela S/R

\section{Stellar parameters}
\label{sec:stellar_parameters}
\subsection{Atmospheric parameters}
\label{sec:atmparam}

We determined the atmospheric parameters of the stars in our sample using the classical spectroscopic method based on the excitation and ionization equilibria of Fe I and Fe II lines. We used ARESv2 (\citealt{sousa2007new}; \citealt{sousa2015ares}) to automatically measure the equivalent widths of 176 lines from the list of \citet{ghezzi2018retired}. We adopted the following input parameters for ARESv2: \textit{smoothder} = 4, \textit{space} = 3.0, \textit{lineresol} = 0.1, and \textit{miniline} = 5. The \textit{rejt} parameter was set to the measured SNR of the spectra as described in the Section \ref{sec:data}.

\begin{figure}[h!]
    \centering
	\includegraphics[width=0.7\columnwidth]{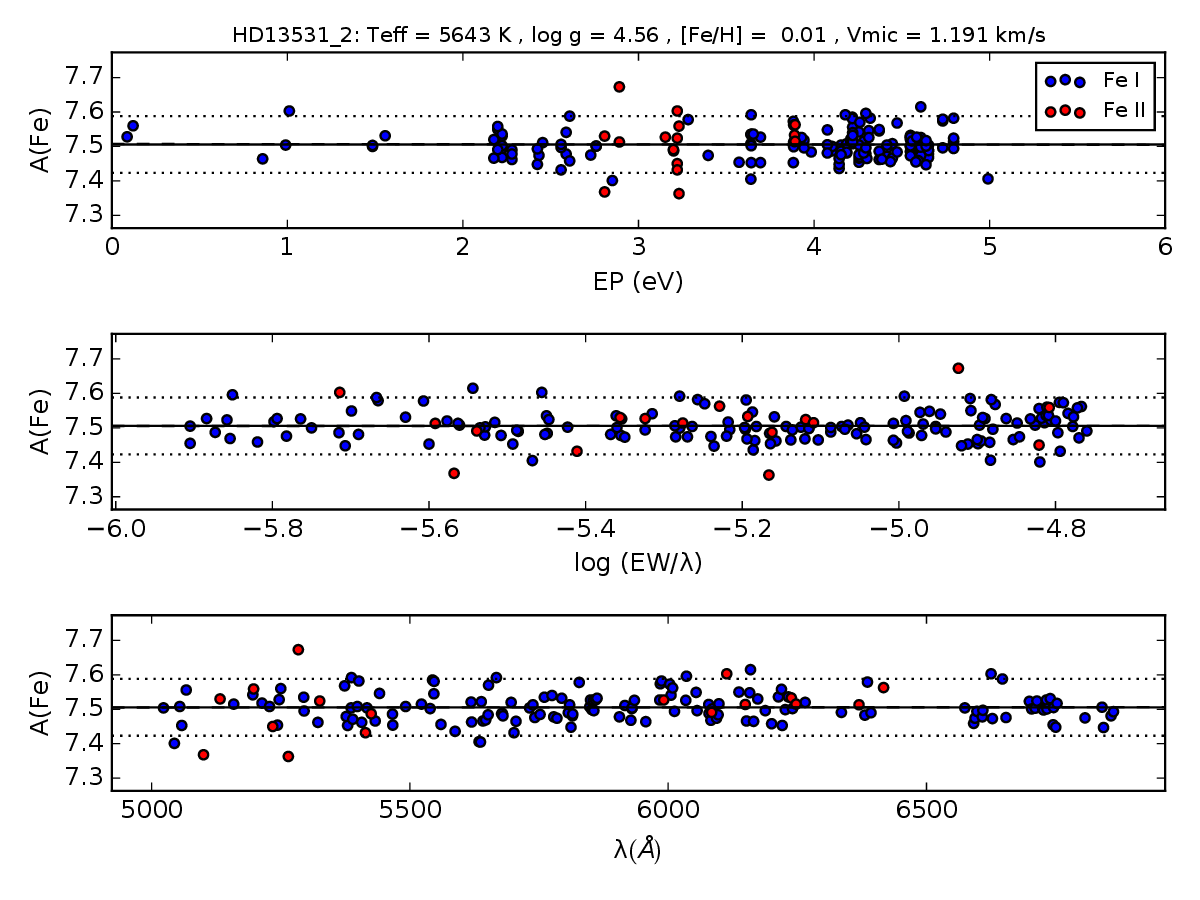}
    \caption{Final result of the atmospheric parameters iteration for the ZAMS candidate HD 13531, using the EW measurements from a ESPaDOnS spectrum (SNR = 310). Upper panel: No trend between the determined A(Fe I) abundances and the excitation potential. Middle panel: No trend between the A(Fe I) abundances and the reduced equivalent width log(EW/$\lambda$) of each line. Lower panel: determined abundances as a function of the wavelength of each spectral line. The blue dots represent the Fe I abundances, while the red dots represent the Fe II abundances. The solid lines show the mean A(Fe) abundance. The dashed lines (which overlaps with the solid line because convergence was achieved) show the linear fits performed. The dotted lines show 2$\sigma$ limits around the mean.}
    \label{fig:example_convergence}
\end{figure}

We determined the atmospheric parameters ($\bm{T}_{eff}$, log $g$, [Fe/H], $\xi$) of the 18 candidates in our sample using the same pipeline as \citet{ghezzi2018retired, 2021ApJ...920...19G}. The analysis was done in LTE using the 2017 version of MOOG\footnote{https://www.as.utexas.edu/~chris/moog.html} \citep{ Sneden1973} along with the ATLAS9 ODFNEW grid \citep{2003IAUS..210P.A20C} to generate interpolated model atmospheres. The final parameters for each star are obtained through a non-differential iterative process that has to simultaneously meet four criteria based on the excitation and ionization equilibria of Fe I and Fe II spectral lines. These criteria are: no correlation between A(Fe I) and the excitation potential $\chi$ (excitation equilibrium) and between A(Fe I) and the reduced equivalent width log(EW/$\lambda$); same average values of A(Fe I) and A(Fe II) (ionization equilibrium); and same value for the metallicity in the input
model atmosphere and the output result from MOOG. This process is repeated by changing the input values of $\bm{T}_{eff}$, $\xi$, log \textit{g}, [Fe/H], respectively, until convergence of the four criteria is achieved. The last step of the automated process is to perform a round of 2$\sigma$ clipping to remove outliers, i.e., lines with corresponding abundances that present large deviations from the average, after that, the entire iteration process is repeated. An example of the final convergence can be seen in the Figure \ref{fig:example_convergence} for the ZAMS candidate HD 13531. The uncertainties of the atmospheric parameters were estimated following the procedures of \citet{ghezzi2018retired, 2021ApJ...920...19G}.

\begin{figure*} % Figura de largura total
    \centering
    \includegraphics[width=0.45\textwidth]{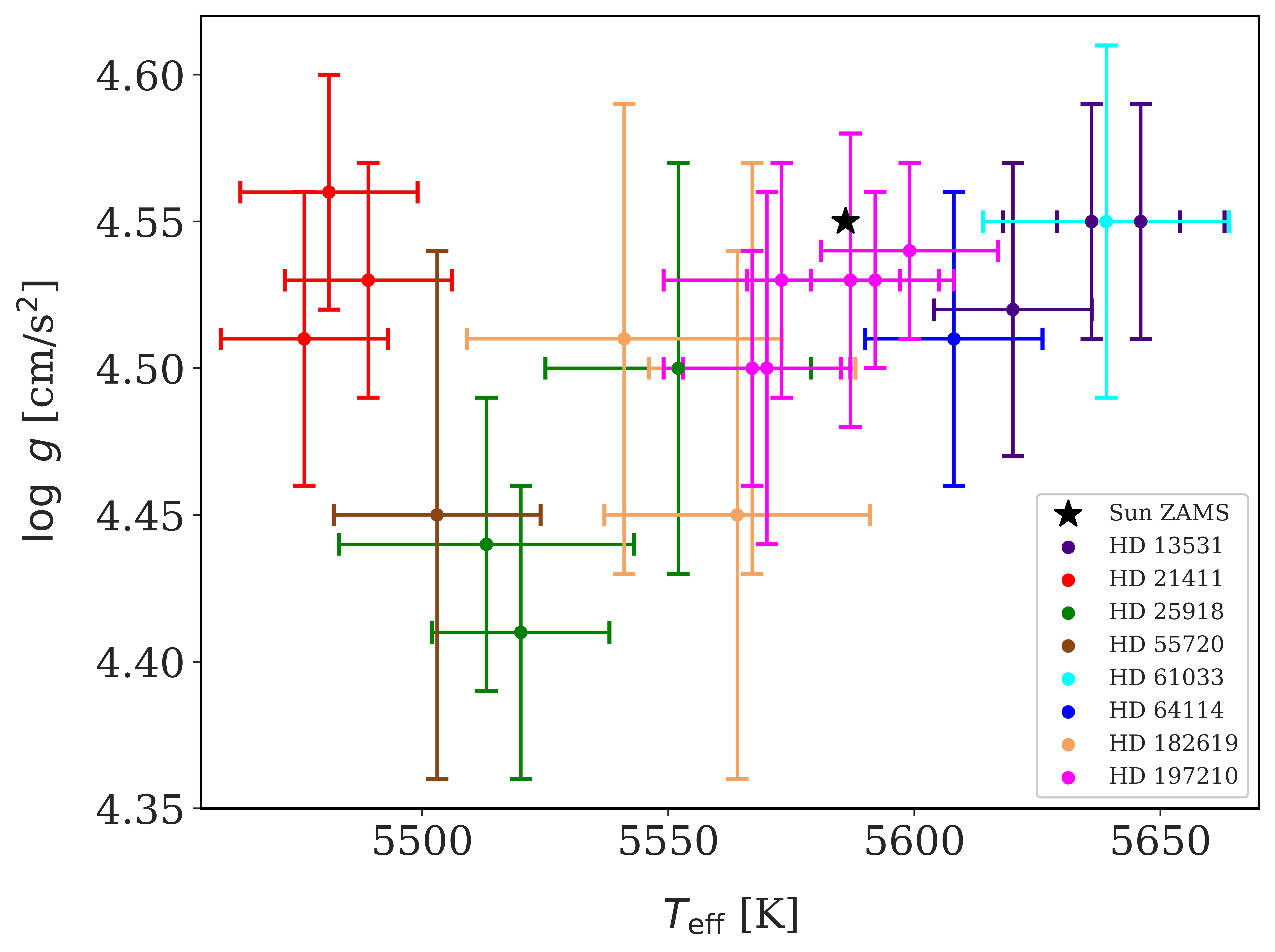} % Ajuste o caminho e o tamanho
    \includegraphics[width=0.45\textwidth]{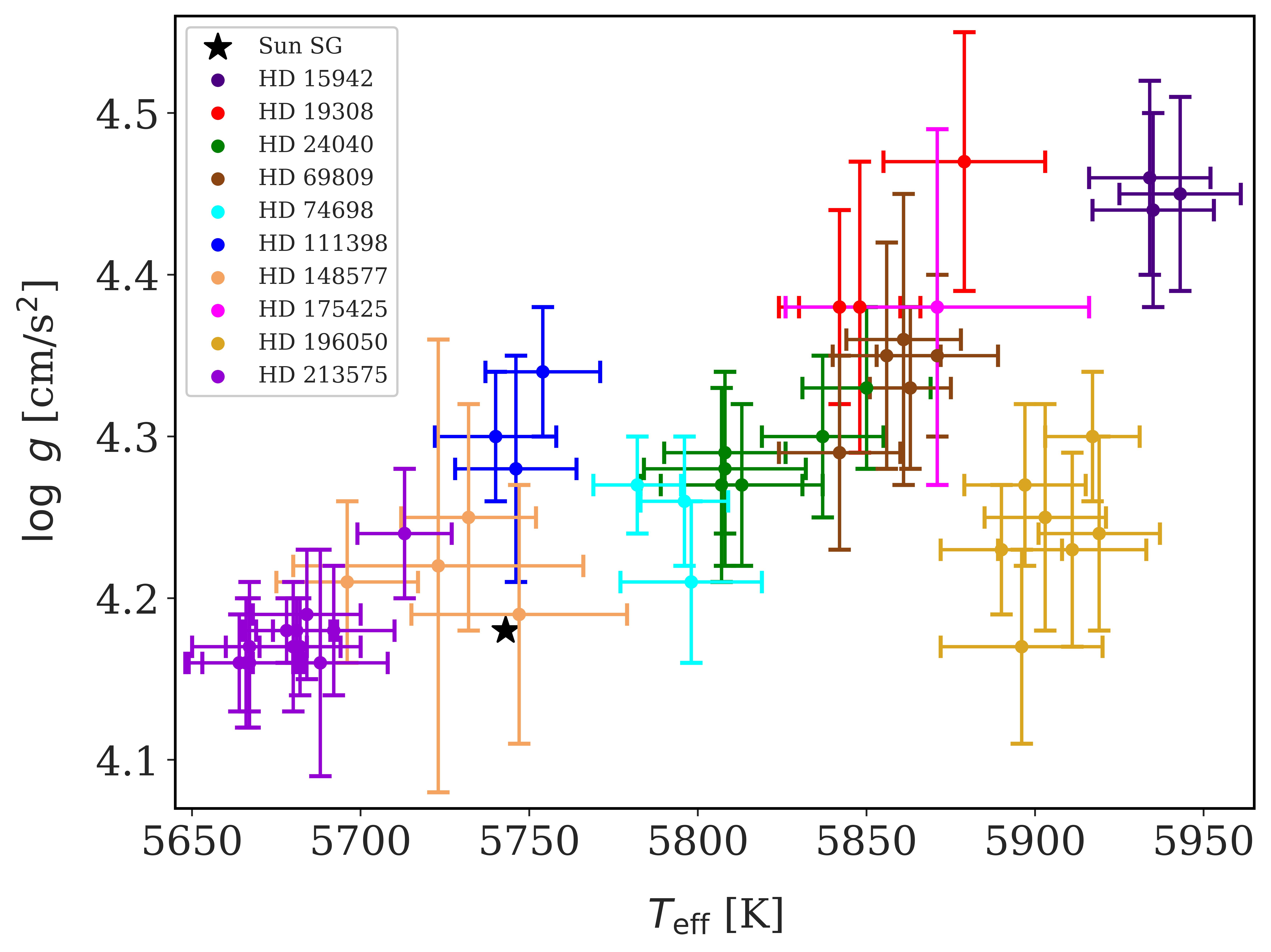} % Ajuste o caminho e o tamanho
    \caption{Effective temperatures $\bm{T}_{eff}$ and surface gravities $\text{log}$ $g$ for all candidates. We present the ZAMS candidates in the left panel and the SG candidates in the right panel. Each unique color represents one of the candidates of the sample. The different values for same colored points represent the results obtained for each spectrum of the same star. The error bars show the respective uncertainties of each atmospheric parameter. The black stars represent the reference atmospheric parameters of the Sun at ZAMS and SG stages respectively.}
    \label{fig:example_ZAMS_SG_Teff_Logg}
\end{figure*}

The derived atmospheric parameters from each spectrum of our sample stars are shown in the Figure \ref{fig:example_ZAMS_SG_Teff_Logg}, where the black stars represent the expected atmospheric parameters of the Sun during the respective evolutionary stage. The final atmospheric parameters and their associated uncertainties can be seen in Table \ref{table_param_atm}. As mentioned before, this analysis was performed for all spectra of our candidate stars. The internal agreement between each observation of the candidates is excellent and their estimated parameters values agree with the mean value within a 2$\sigma$ interval for every candidate.

\begin{table}[h!]
    \centering
    \caption{Final atmospheric parameters of all sample candidates. In order, the columns present: the star identification, its effective temperature, logarithm of the surface gravity, metallicity and the microturbulence velocity.}
    \begin{tabular}{lcccc}
        \hline
	Star          & $\bm{T}_{eff}$   & \text{log} $g$   & $[Fe/H]$  & $\xi$   \\
                    &(K)     & $(cms^{-2})$                &(dex)              &$(km s^{-1})$\\
        \hline
        Sun ZAMS & 5586 & 4.53 & 0.00 &  -- \\
        HD 13531 & 5634 $\pm$ 17 & 4.54 $\pm$ 0.04 & -0.01 $\pm$ 0.02 & 1.22 $\pm$ 0.04 \\
        HD 21411 & 5482 $\pm$ 17 & 4.53 $\pm$ 0.04 & -0.22 $\pm$ 0.01 & 0.85 $\pm$ 0.05 \\
        HD 25918 & 5528 $\pm$ 25 & 4.45 $\pm$ 0.06 & -0.05 $\pm$ 0.02 & 0.77 $\pm$ 0.05 \\
        HD 55720 & 5503 $\pm$ 21 & 4.45 $\pm$ 0.09 & -0.28 $\pm$ 0.02 & 0.93 $\pm$ 0.04 \\
        HD 61033 & 5639 $\pm$ 25 & 4.55 $\pm$ 0.06 & 0.02  $\pm$ 0.02 & 1.12 $\pm$ 0.05 \\
        HD 64114 & 5608 $\pm$ 18 & 4.51 $\pm$ 0.05 & 0.01  $\pm$ 0.01 & 0.96 $\pm$ 0.03 \\
        HD 182619 & 5557 $\pm$ 27 & 4.49 $\pm$ 0.08 & -0.07 $\pm$ 0.02 & 0.82 $\pm$ 0.06  \\
        HD 197210 & 5581 $\pm$ 19 & 4.52 $\pm$ 0.04 & 0.01  $\pm$ 0.01 & 0.88 $\pm$ 0.04  \\
        \hline
        Sun SG & 5743 & 4.18 & 0.00 &  -- \\
        HD 15942 & 5934 $\pm$ 11 & 4.45 $\pm$ 0.06 & 0.43 $\pm$ 0.01 & 1.13 $\pm$ 0.03 \\
        HD 19308 & 5856 $\pm$ 20 & 4.41 $\pm$ 0.08 & 0.16 $\pm$ 0.02 & 1.02 $\pm$ 0.03 \\
        HD 24040 & 5821 $\pm$ 21 & 4.29 $\pm$ 0.05 & 0.20 $\pm$ 0.02 & 1.10 $\pm$ 0.04  \\
        HD 69809 & 5859 $\pm$ 16 & 4.34 $\pm$ 0.06 & 0.28 $\pm$ 0.01 & 1.13 $\pm$ 0.03 \\
        HD 74698 & 5792 $\pm$ 16 & 4.25 $\pm$ 0.04 & 0.12 $\pm$ 0.01 & 1.08 $\pm$ 0.03 \\
        HD 111398 & 5747 $\pm$ 18 & 4.31 $\pm$ 0.05 & 0.08 $\pm$ 0.01 & 1.04 $\pm$ 0.03 \\
        HD 148577 & 5725 $\pm$ 29 & 4.22 $\pm$ 0.09 & -0.03 $\pm$ 0.02 & 1.08 $\pm$ 0.05 \\
        HD 175425 & 5871 $\pm$ 45 & 4.38 $\pm$ 0.11 & 0.15 $\pm$ 0.03 & 1.13 $\pm$ 0.08 \\
        HD 196050 & 5905 $\pm$ 19 & 4.24 $\pm$ 0.05 & 0.25 $\pm$ 0.01 & 1.21 $\pm$ 0.03 \\
        HD 213575 & 5680 $\pm$ 16 & 4.18 $\pm$ 0.04 & -0.14 $\pm$ 0.01 & 1.09 $\pm$ 0.03 \\
        \hline
        Present Sun & 5777 & 4.44 & 0.00 & 1.00 \\
        18 Sco  & 5811 $\pm$ 17 & 4.47 $\pm$ 0.06 & 0.06 $\pm$ 0.01 & 1.03 $\pm$ 0.03 \\
        \hline
    \end{tabular}
\label{table_param_atm}
\end{table}

In the end of the process, we took the average values as the final atmospheric parameters for a given star. We estimated the final errors for each parameter by taking the average of the error values. We applied this procedure because the propagated standard deviation was too small and would underestimate the expected uncertainties. For example, in the case of the Young Sun candidates, the mean standard deviation for stars with at least three spectra was $\approx$ 21 K, 0.05 dex and 0.02 dex for $T_{\mathrm{eff}}$, \text{log} $g$ and [Fe/H], respectively. For the SG candidates, the mean standard deviation for stars with at least three spectra was $\approx$ 18 K, 0.06 dex and 0.01 dex. These results reinforce the internal consistency of our atmospheric parameters, even though spectra from different instruments were used (see Section \ref{sec:data}). This consistency is also illustrated in Figure~\ref{fig:example_ZAMS_SG_Teff_Logg}, where the points representing the parameters for each spectrum are closely grouped for all stars.

%In the end of the process we took the average values as the final atmospheric parameters for a given star. We estimated the final errors for each parameter by taking the average of the error values. We applied this procedure because the propagated standard deviation was too small and would not adequately reflect the expected uncertainties. \textbf{For example, in the case of the Young Sun candidates, the mean standard deviation for stars with at least three spectra was $\approx$ 21K, 0.05$(cms^{-2})$ and 0.02 dex for $T_{\mathrm{eff}}$, \text{log} $g$ and [Fe/H], respectively. For the SG candidates, the mean standard deviation for stars with at least three spectra was $\approx$ 18K, 0.06 $(cms^{-2})$ and 0.01 dex. These results reinforce the internal consistency of our analysis, as illustrated in Figure~\ref{fig:example_ZAMS_SG_Teff_Logg}, where the points representing the parameters for each spectrum are closely grouped for individual stars.} 

%For example, in the case of the effective temperatures ($\bm{T}_{eff}$), the mean standard deviation for stars with at least three spectra was $\approx$13 K, which reinforces the internal consistency of our results.

We chose the star HD 146233 (18 Sco) as a control star in our sample in order to check the precision of our analysis during the project. Our derived atmospheric parameters for HD 146233 have a good agreement with previous works (e.g., \citealt{de1997hr}; \citealt{bazot2011radius}; \citealt{de2014photometric}; \citealt{melendez201418}; \citealt{heiter2015gaia}) as shown in Table \ref{table_18Sco}.

\begin{table*}[h!]
    \centering
    \caption{Final atmospheric parameters of the solar twin HD 146233 (18 Sco) and comparison with previous works in the literature. In order, the columns present: the reference article, effective temperature, logarithm of the surface gravity, metallicity and the microturbulence velocity.}
    \begin{tabular}{lllll}
    \hline
                  & $\bm{T}_{eff}$   & $\text{log}$ $g$   & $[Fe/H]$  & $\xi$\\
                    &(K)     & $(cms^{-2})$                &(dex)              &$(km s^{-1})$\\
        \hline
        %Sun & 5777 & 4.44 & 0.00 & 1.00 \\
        This Work  & 5811 $\pm$ 17 & 4.47 $\pm$ 0.06 & 0.06 $\pm$ 0.01 & 1.03 $\pm$ 0.03\\
        Bazot et al. (2011)  & 5813 $\pm$ 21 & 4.45 $\pm$ 0.02 & 0.04 $\pm$ 0.01 & -\\
        Porto de Mello et al. (2014) & 5795 $\pm$ 30 & 4.42 $\pm$ 0.05 & -0.03 $\pm$ 0.04 & -\\
        Meléndez et al. (2014) & 5823 $\pm$ 6 & 4.45 $\pm$ 0.02 & 0.05 $\pm$ 0.01 & -\\
        Gaia Benchmark Stars (2015) & 5786 $\pm$ 30 & 4.39 $\pm$ 0.07 & 0.03 $\pm$ 0.01 & 1.20 $\pm$ 0.20\\
        
        \hline
    \end{tabular}
\label{table_18Sco}
\end{table*}

We note that the choice of an analysis in 1D LTE does not affect the selection of the best candidates because the theoretical points ZAMS and SG both have parameters very close to the solar ones. Therefore, the 3D NLTE corrections for the parameter space studied in this work are $A(X)$ $\lesssim$ 0.03 dex (e.g., \citealt{2022A&A...668A..68A}; \citealt{2024ARA&A..62..475L}), a value that is within our uncertainties. Moreover, we compare each candidate's parameters with those from the corresponding theoretical point, making possible systematic effects negligible. 

%\textbf{We expect small differences associated with systematic errors arising from the adopted approximations (MLT, plane-parallel geometry, and LTE), which depend on the different methods and models employed. However, since our sample was selected around a theoretical point of solar evolution, our candidates are affected to a lesser extent. \citet{2022A&A...668A..68A} and \citet{2024ARA&A..62..475L} mention that abundance corrections relative to solar values are less severe ($A_{approx}$ $\approx$ 0.03 dex for several chemical species). In this context, these well-known deviations are not expected to compromise the significance of the candidates highlighted throughout this project. In this sense, we expect the influence of these effects to be negligible within our sample because we are comparing the stars and the theoretical points within the same evolutionary stage, that is, with very similar parameters.}

Atomic diffusion could, in principle, also affect the selection of the best candidates to represent the Sun throughout the main sequence because it can lower the surface iron abundance of solar-type stars due to gravitational settling by 0.1 dex (\citealt{2017ApJ...840...99D}; \citealt{2025A&ARv..33....3G}). However, we note that no candidate was excluded solely because of its metallicity (see Section \ref{best_candidates}), but rather due to a combination of its physical parameters (primarily age, which ruled out most of the SG candidates).

%\textbf{It is important to emphasize that the effect of atomic diffusion should be taken into account when considering the metallicity of the SG candidate stars in our project, as studies in the literature show that this effect can reach values of up to -0.1 dex in the surface iron abundance for solar-type stars due to gravitational settling (\citealt{2017ApJ...840...99D}; \citealt{2025A&ARv..33....3G}). Nevertheless, within the context of the SOL Project results, this effect did not influence the final highlighted SG candidates, since candidates at the same evolutionary stage would be affected in a similar manner. Furthermore, no candidate was excluded solely because of its metallicity (see Section \ref{best_candidates}), but rather due to a combination of its physical parameters (primarily age, which ruled out most of the SG candidates)}.

\subsection{Chemical abundances}
\label{sec:abundances}

Using the atmospheric parameters from Section \ref{sec:atmparam}, we determined abundances of the following elements: carbon (C), oxygen (O), sodium (Na), magnesium (Mg), aluminum (Al), silicon (Si), calcium (Ca), titanium (Ti), nickel (Ni), chromium (Cr) and zinc (Zn). We used the code ARESv2 \citep{sousa2007new,sousa2015ares} to automatically measure EWs for a list of lines with solar $\text{log \textit{gf}}$ values that will be described in a future paper (Ghezzi et al., in prep.). In total, we measured 166 absorption lines from 11 different elements, spanning wavelengths from 4400 $\AA$ to 7800 \AA; a spectral range covered by the majority of instruments used throughout this work. The determination of abundances was performed using the 2017 version of MOOG and the \textit{abfind} task.

The abundance uncertainties for each element in each spectrum were estimated as follows: taking $\bm{T}_{eff}$ as an example, with its associated uncertainty $\sigma_{\bm{T}_{eff}}$ (as determined in Section~\ref{sec:atmparam}), a new model atmosphere was computed using $\bm{T}_{eff}$ - $\sigma_{\bm{T}_{eff}}$. All elemental abundances were recalculated, and the difference $A(\mathrm{X})_{\mathrm{new}} - A(\mathrm{X})_{\mathrm{final}}$ was derived. The same procedure was repeated for $\bm{T}_{eff}$ + $\sigma_{\bm{T}_{eff}}$, resulting in a pair of abundance deviations for each element. The maximum of the two values was adopted as the uncertainty due to $\bm{T}_{eff}$. This process was repeated for the other atmospheric parameters: $\text{log}$ $g$, [Fe/H], and $\xi$. Finally, the total uncertainty in the abundance of each element was computed as the quadratic sum of five terms: the four contributions from the atmospheric parameters ($\bm{T}_{eff}$, $\text{log}$ $g$, [Fe/H], and $\xi$) and the standard deviation of the mean of the final abundance measurements. This latter contribution was only considered for elements with two or more lines.

For the determination of O abundance, we utilized the \ion{O}{1} triplet at 777 nm. However, it is well established in the literature that NLTE effects influence O abundance determinations in solar-type stars, on the order of a few tenths of a dex \citep{KISELMAN2001559}. Briefly, 1D LTE analyzes of these \ion{O}{1} lines tend to yield overestimated O abundances. To account for these NLTE effects, we applied the corrections derived by \cite{2019A&A...630A.104A}.

\begin{figure}[h!]
    \centering
	\includegraphics[width=0.7\columnwidth]{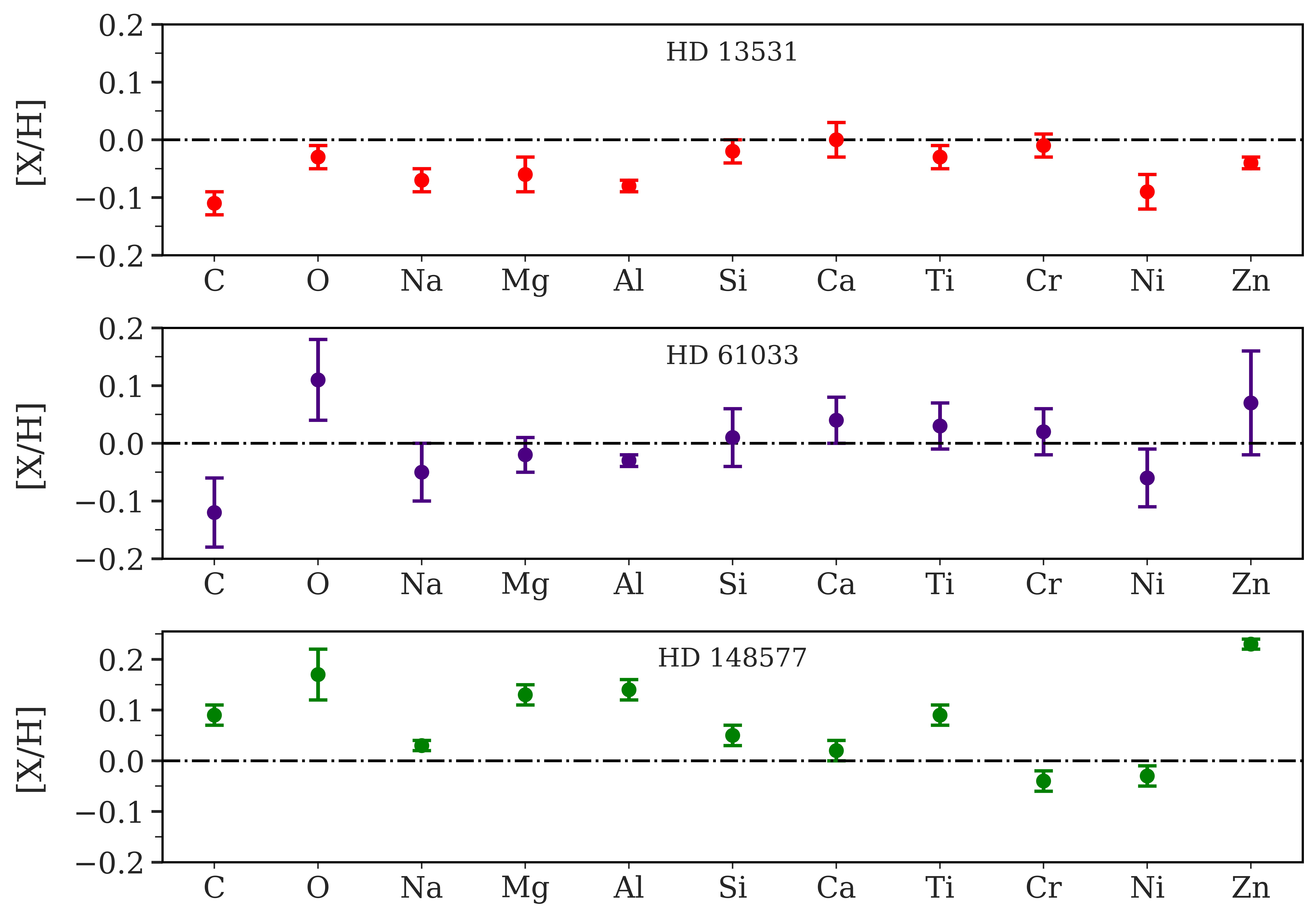}
    \caption{Chemical abundances for HD 13531 (ZAMS candidate; top panel), HD 61033 (ZAMS candidate; middle panel) and HD 148577 (SG candidate; lower panel). The dotted-dashed line marks the solar reference ([X/H] = 0.00).}
    \label{fig:abundances_x_h}
\end{figure}

We averaged the mean abundance of each element to obtain its final chemical abundance. The determined absolute abundances were subtracted from the solar reference abundances adopted from Table 1 of \citet{asplund2009chemical}. Therefore, the abundance values presented here follow the notation [X/H], where X represents the element, and the abundance value is given in dex with its associated uncertainty, as presented in Table \ref{tab:abundancias}. Figure \ref{fig:abundances_x_h} shows the final chemical abundances of the best candidates in the sample (as it will be shown throughout the paper). The candidate HD 61033 shows an abundance pattern consistent with the solar reference, whereas HD 13531 is slightly more metal-poor than the Sun and HD 148577 is slightly more metal-rich. 
We performed linear fits for the abundances as a function of their condensation temperatures (taken from \citealt{Lodders_2003}), but found no statistically significant correlations for any of the candidates.

%\textbf{We tested the possibility of significant abundance trends with condensation temperature, but we found no indication of such behavior for any of the candidates}.

Note that the abundances of Ti and Cr are the weighted average between the values obtained from different ionization stages. We compared the abundances determined using Ti I/Ti II and Cr I/Cr II lines using weighted linear fits. The slopes were 1.095 $\pm$ 0.036 and 0.950 $\pm$ 0.015, respectively, with $\text{R}^2$ of 0.983 and 0.996. Additionally, the mean abundance difference (X I - X II) obtained between the ionized stages is -0.006 $\pm$ 0.007 dex for Ti, and -0.014 $\pm$ 0.005 dex for Cr. These results strengthen our confidence in the procedures adopted, as well as in the accuracy of the atmospheric parameter determination carried out in Section \ref{sec:atmparam}. The abundances of Ti and Cr presented in Table \ref{tab:abundancias} are also the weighted mean of the ionization stages. 

%\begin{landscape}
\begin{longrotatetable}
\begin{table*}
\centering
\caption{Final chemical abundances ([X/H]) for all sample stars. The columns present the stellar ID followed by the abundance measurements (in dex) and their uncertainties for each chemical element. Weighted averages are shown for Cr and Ti.}
\label{tab:abundancias}
\begin{tabular}{lccccccccccc}
\hline
Star & [C/H] & [O/H] & [Na/H] & [Mg/H] & [Al/H] & [Si/H] & [Ca/H] & [Ti/H] & [Cr/H] & [Ni/H] & [Zn/H] \\
\hline
HD 13531 & $-0.11 \pm 0.02$ & $-0.03 \pm 0.02$ & $-0.07 \pm 0.02$ & $-0.06 \pm 0.03$ & $-0.08 \pm 0.01$ & $-0.02 \pm 0.02$ & $0.00 \pm 0.03$ & $-0.03 \pm 0.02$ & $-0.01 \pm 0.02$ & $-0.09 \pm 0.03$ & $-0.04 \pm 0.01$ \\
\hline
HD 21411 & $-0.21 \pm 0.02$ & - & $-0.29 \pm 0.01$ & $-0.19 \pm 0.03$ & $-0.23 \pm 0.02$ & $-0.21 \pm 0.02$ & $-0.17 \pm 0.02$ & $-0.18 \pm 0.02$ & $-0.20 \pm 0.02$ & $-0.28 \pm 0.02$ & $-0.23 \pm 0.02$ \\
\hline
HD 25918 & $-0.14 \pm 0.02$ & - & $-0.22 \pm 0.02$ & $-0.04 \pm 0.04$ & $0.01 \pm 0.02$ & $-0.07 \pm 0.02$ & $-0.04 \pm 0.03$ & $-0.01 \pm 0.02$ & $-0.05 \pm 0.02$ & $-0.11 \pm 0.02$ & $-0.06 \pm 0.01$ \\
\hline
HD 55720 & $-0.19 \pm 0.04$ & $-0.05 \pm 0.05$ & $-0.26 \pm 0.05$ & $-0.07 \pm 0.06$ & $-0.11 \pm 0.05$ & $-0.16 \pm 0.03$ & $-0.17 \pm 0.05$ & $-0.09 \pm 0.03$ & $-0.28 \pm 0.04$ & $-0.28 \pm 0.04$ & $-0.12 \pm 0.06$ \\
\hline
HD 61033 & $-0.12 \pm 0.06$ & $0.11 \pm 0.07$ & $-0.05 \pm 0.05$ & $-0.02 \pm 0.03$ & $-0.03 \pm 0.01$ & $0.01 \pm 0.05$ & $0.04 \pm 0.04$ & $0.03 \pm 0.04$ & $0.02 \pm 0.04$ & $-0.06 \pm 0.05$ & $0.07 \pm 0.09$ \\
\hline
HD 64114 & $-0.08 \pm 0.04$ & $-0.06 \pm 0.04$ & $-0.02 \pm 0.06$ & $0.06 \pm 0.07$ & $-0.01 \pm 0.01$ & $0.02 \pm 0.03$ & $0.02 \pm 0.04$ & $0.04 \pm 0.03$ & $0.01 \pm 0.03$ & $-0.02 \pm 0.04$ & $0.03 \pm 0.06$ \\
\hline
HD 182619 & $-0.11 \pm 0.04$ & - & $-0.12 \pm 0.02$ & $-0.06 \pm 0.04$ & $-0.05 \pm 0.02$ & $-0.05 \pm 0.02$ & $-0.07 \pm 0.03$ & $-0.04 \pm 0.02$ & $-0.07 \pm 0.02$ & $-0.09 \pm 0.02$ & $-0.02 \pm 0.02$ \\
\hline
HD 197210 & $-0.07 \pm 0.01$ & - & $-0.04 \pm 0.02$ & $0.01 \pm 0.02$ & $0.02 \pm 0.02$ & $0.00 \pm 0.01$ & $0.02 \pm 0.01$ & $0.02 \pm 0.01$ & $0.02 \pm 0.01$ & $-0.02 \pm 0.01$ & $0.03 \pm 0.01$ \\
\hline
HD 15942 & $0.26 \pm 0.02$ & - & $0.49 \pm 0.02$ & $0.43 \pm 0.03$ & $0.42 \pm 0.01$ & $0.46 \pm 0.02$ & $0.40 \pm 0.02$ & $0.45 \pm 0.02$ & $0.44 \pm 0.02$ & $0.47 \pm 0.02$ & $0.56 \pm 0.01$ \\
\hline
HD 19308 & $0.12 \pm 0.02$ & - & $0.22 \pm 0.02$ & $0.19 \pm 0.02$ & $0.19 \pm 0.01$ & $0.18 \pm 0.02$ & $0.10 \pm 0.02$ & $0.15 \pm 0.02$ & $0.16 \pm 0.02$ & $0.18 \pm 0.02$ & $0.16 \pm 0.02$ \\
\hline
HD 24040 & $0.14 \pm 0.02$ & $0.21 \pm 0.01$ & $0.25 \pm 0.01$ & $0.20 \pm 0.01$ & $0.27 \pm 0.01$ & $0.23 \pm 0.02$ & $0.20 \pm 0.02$ & $0.24 \pm 0.02$ & $0.20 \pm 0.02$ & $0.23 \pm 0.02$ & $0.29 \pm 0.01$ \\
\hline
HD 69809 & $0.26 \pm 0.02$ & $0.15 \pm 0.02$ & $0.43 \pm 0.01$ & $0.36 \pm 0.02$ & $0.34 \pm 0.01$ & $0.32 \pm 0.02$ & $0.24 \pm 0.02$ & $0.30 \pm 0.02$ & $0.29 \pm 0.02$ & $0.34 \pm 0.01$ & $0.23 \pm 0.01$ \\
\hline
HD 74698 & $0.09 \pm 0.02$ & - & $0.07 \pm 0.02$ & $0.17 \pm 0.03$ & $0.16 \pm 0.01$ & $0.13 \pm 0.02$ & $0.12 \pm 0.02$ & $0.14 \pm 0.02$ & $0.12 \pm 0.02$ & $0.10 \pm 0.02$ & $0.21 \pm 0.01$ \\
\hline
HD 111398 & $0.06 \pm 0.02$ & - & $0.06 \pm 0.02$ & $0.18 \pm 0.02$ & $0.19 \pm 0.02$ & $0.12 \pm 0.02$ & $0.09 \pm 0.03$ & $0.15 \pm 0.02$ & $0.09 \pm 0.02$ & $0.08 \pm 0.02$ & $0.23 \pm 0.01$ \\
\hline
HD 148577 & $0.09 \pm 0.02$ & $0.17 \pm 0.05$ & $0.03 \pm 0.01$ & $0.13 \pm 0.02$ & $0.14 \pm 0.02$ & $0.05 \pm 0.02$ & $0.02 \pm 0.02$ & $0.09 \pm 0.02$ & $-0.04 \pm 0.02$ & $-0.03 \pm 0.02$ & $0.23 \pm 0.01$ \\
\hline
HD 175425 & $0.11 \pm 0.09$ & - & $0.11 \pm 0.06$ & $0.15 \pm 0.04$ & $0.26 \pm 0.03$ & $0.17 \pm 0.08$ & $0.07 \pm 0.05$ & $0.20 \pm 0.07$ & $0.08 \pm 0.05$ & $0.14 \pm 0.08$ & $0.19 \pm 0.02$\\
\hline
HD 196050 & $0.21 \pm 0.01$ & $0.08 \pm 0.02$ & $0.39 \pm 0.01$ & $0.27 \pm 0.01$ & $0.30 \pm 0.01$ & $0.29 \pm 0.01$ & $0.23 \pm 0.02$ & $0.29 \pm 0.01$ & $0.26 \pm 0.01$ & $0.30 \pm 0.01$ & $0.40 \pm 0.01$ \\
\hline
HD 213575 & $0.01 \pm 0.01$ & - & $-0.08 \pm 0.01$ & $0.06 \pm 0.01$ & $0.07 \pm 0.01$ & $-0.01 \pm 0.01$ & $-0.04 \pm 0.01$ & $0.04 \pm 0.01$ & $-0.13 \pm 0.01$ & $-0.13 \pm 0.01$ & $-0.02 \pm 0.01$ \\
\hline
\end{tabular}
\end{table*}
\end{longrotatetable}
%\end{landscape}

\subsection{Evolutionary parameters}
\label{sec:evolparam}

We obtained the evolutionary parameters (mass, radius, luminosity, and age) for every target star in our sample. This task was performed using the \textit{isoclassify}\footnote{https://github.com/danxhuber/isoclassify} code (\citealt{huber2017asteroseismology}; \citealt{berger2020gaia}) which compares observational parameters with grids of isochrones. We opted for the grid modeling mode of the code to derive the stellar parameters (for example, radius, mass, luminosity and age) which performs a direct integration of isochrones with a given set of observables as input (photometry, spectroscopy and parallax), generating posterior distributions for each stellar parameter. In our analysis of the sample, we chose to use the set of isochrones from the MESA Isochrones and Stellar Tracks (MIST) project \citep{2016ApJS..222....8D, 2016ApJ...823..102C}. For the extinction calculation we chose the "all sky" option, which combines the reddening maps from \citet{bovy2016galactic} and \textit{Bayestar19} (\citealt{green20193d}). However, it is important to note that we should not expect any major contribution from reddening and extinction in the photometry of our sample due to their proximity; our most distant candidate (HD 13531), excluding the control star, is located 26.13 pc away, as determined based on its parallax (38.27 \textit{mas}). In fact, it was only possible to estimate this correction for 5 candidates, with the highest value $A_{V} \approx 0.04$ (extinction in the \textit{V} magnitude), almost negligible, as expected.

We also chose to provide the input parameters $\bm{T}_{eff}$, \text{log} \textit{g}, [Fe/H], \textit{V} magnitude, celestial equatorial coordinates (\textit{RA}, \textit{DEC}) and parallax $\pi$, in addition to their respective uncertainties. The effective temperatures, the logarithm of the surface gravities and metallicities were determined spectroscopically in this work as described earlier (see Section \ref{sec:atmparam}). The \textit{RA} and \textit{DEC} coordinates and the parallaxes of the candidates, with their associated uncertainties, were collected from the second Data Release (DR2) of Gaia \citep{gaia2018gaia}. The \textit{V} magnitudes were calculated using the following formula: $V = V_{T} - [0.090(B_{T} - V_{T})]$; the values of B$_{T}$ and V$_{T}$ were obtained from the \textit{Tycho-2} catalog \citep{2000A&A...355L..27H}, for consistency with the data used during the selection of the sample, as described in Section \ref{sec:sample_selection_data}. Uncertainties for the \textit{V} magnitudes were determined through error propagation.

\begin{figure}
    \centering
	\includegraphics[width=0.7\columnwidth]{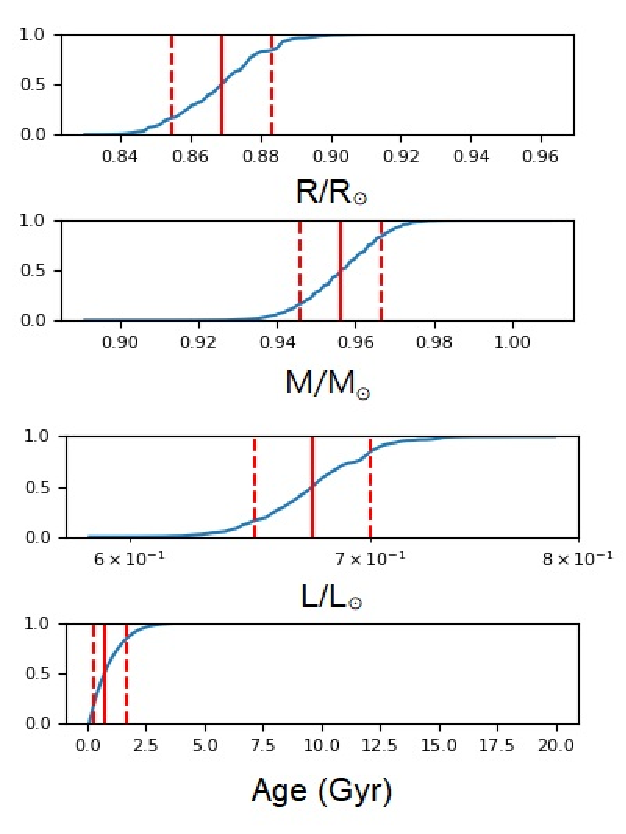}
    \caption{The output of the isoclassify code for the ZAMS candidate HD 61033. From the upper to lower panels we have the fits for the radius, mass, luminosity and age, respectively. The blue line represents the cumulative distribution function, while the solid red line represents the best value fitted. The dashed red lines represent uncertainties for the best value.}
    \label{fig:example_fit}
\end{figure}

\begin{figure}
    \centering
	\includegraphics[width=0.7\columnwidth]{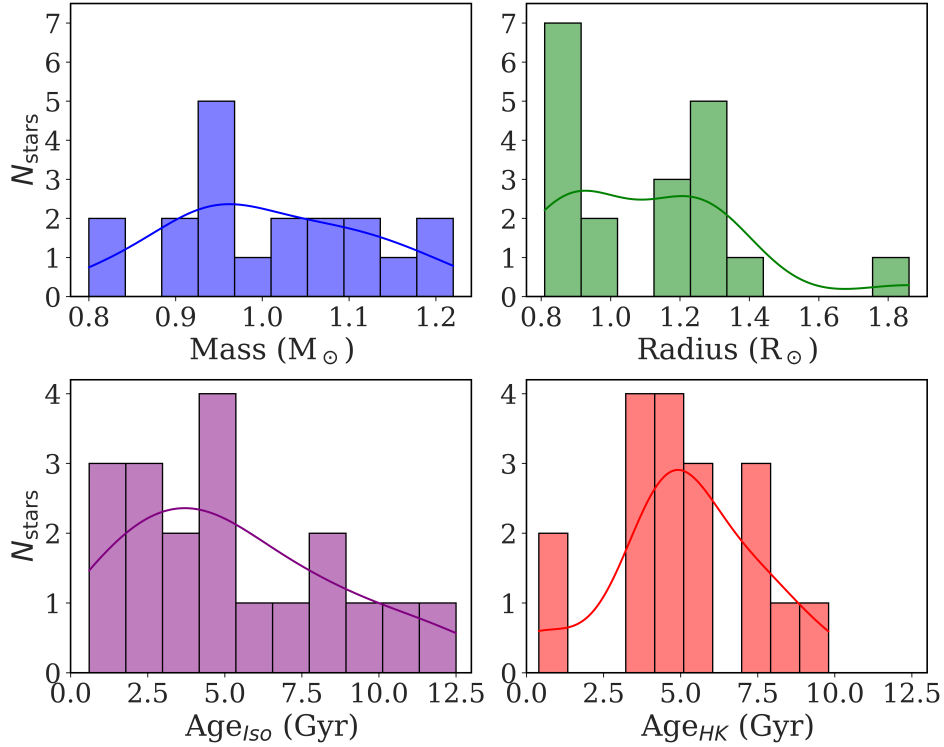}
    \caption{Distribution of stellar parameters for the entire sample. The panels show histograms of stellar mass, radius, and two age estimates (from isochrone fitting, \( \mathrm{Age}_{\mathrm{Iso}} \), and from chromospheric activity, \( \mathrm{Age}_{\mathrm{HK}} \)). The y-axis in all panels represents the number of stars in each bin. A kernel density estimate (KDE) is overplotted to highlight the underlying distribution trends.}
    \label{fig:dist_histograms}
\end{figure}

An example of the cumulative distribution functions (CDFs) for the evolutionary parameters fit for the ZAMS candidate HD 61033 can be seen in Figure \ref{fig:example_fit}. The individual values for the evolutionary parameters of each star in our sample can be seen in the Table \ref{table_result_finais_complete}. The distribution of masses, radii and ages for our sample show that we were capable of identifying interesting candidates to represent the Sun at ZAMS and SG stages as shown in Figure \ref{fig:dist_histograms}. Additionally, we included a Kernel Density Estimation (KDE) in the histograms of Figure \ref{fig:dist_histograms}, where their global peaks are located at 0.96 $M_{\odot}$, 1.21 $R_{\odot}$, 3.71 Gyr for \( \mathrm{Age}_{\mathrm{iso}} \), and 4.94 Gyr for \( \mathrm{Age}_{\mathrm{HK}} \) (see Section \ref{sec:HK_lines}). As can be seen, the candidates HD 13531 and HD 61033 do not have ages consistent with the ZAMS stage; however, they are very close to the parameters expected for young solar analogs, and we will hereafter consider them as such.

%Figure \ref{fig:grafico_final} shows some stars close to the expected reference values of each proposed evolutionary stage, reinforcing the effectiveness our selection method.

%\begin{figure}
%    \centering
%	\includegraphics[width=0.7\columnwidth]{Graph_Lum_Teff.jpg}
%    \caption{HR diagram for all candidates of the sample. The color bar represents the [Fe/H] of each candidate. The colored stars show the reference values for the Sun at different evolutionary stages. The gray dash-dotted and black solid curve represent the Sun’s evolutionary trajectory by \citet{schaller1992new} and PARSEC Stellar Tracks \citep{bressan2012parsec}, respectively, starting at the ZAMS. The diamonds represent the values of the ZAMS candidates, while the dots represent the SG candidates. Our control star 18 Sco is represented by the  triangle.}
%    \label{fig:grafico_final}
%\end{figure}

\subsection{Kinematics parameters}
\label{sec:evol_kine}

We also derived \textit{UVW} components of the galactic space velocities for each star in our sample to check if their kinematics presented the expected behavior according to their evolutionary stage. To accomplish this we used the \textit{gal\textunderscore uvw} code\footnote{\url{https://github.com/segasai/astrolibpy/blob/master/astrolib/gal_uvw.py}} from Astrolibpy package. This routine uses celestial equatorial coordinates (\textit{RA}, \textit{DEC}), proper motion on both coordinates ($\mu_{\alpha}$ and $\mu_{\delta}$), parallax ($\pi$) and radial velocities (\textit{V$_{rad}$}) as input parameters to calculate the galactic components relative to the local standard of rest (LSR). All the input parameters were obtained from the second Data Release (DR2) of Gaia \citep{gaia2018gaia}. We adopted the code option of performing, at the end of the calculation, a correction relative to the solar components on the reference LSR, subtracting the \textit{UVW} solar velocities, obtained by \citet{cocskunouglu2011local}, from the calculated velocities of each target star. 

\begin{figure*} % Figura de largura total
    \centering
    \includegraphics[width=0.45\textwidth]{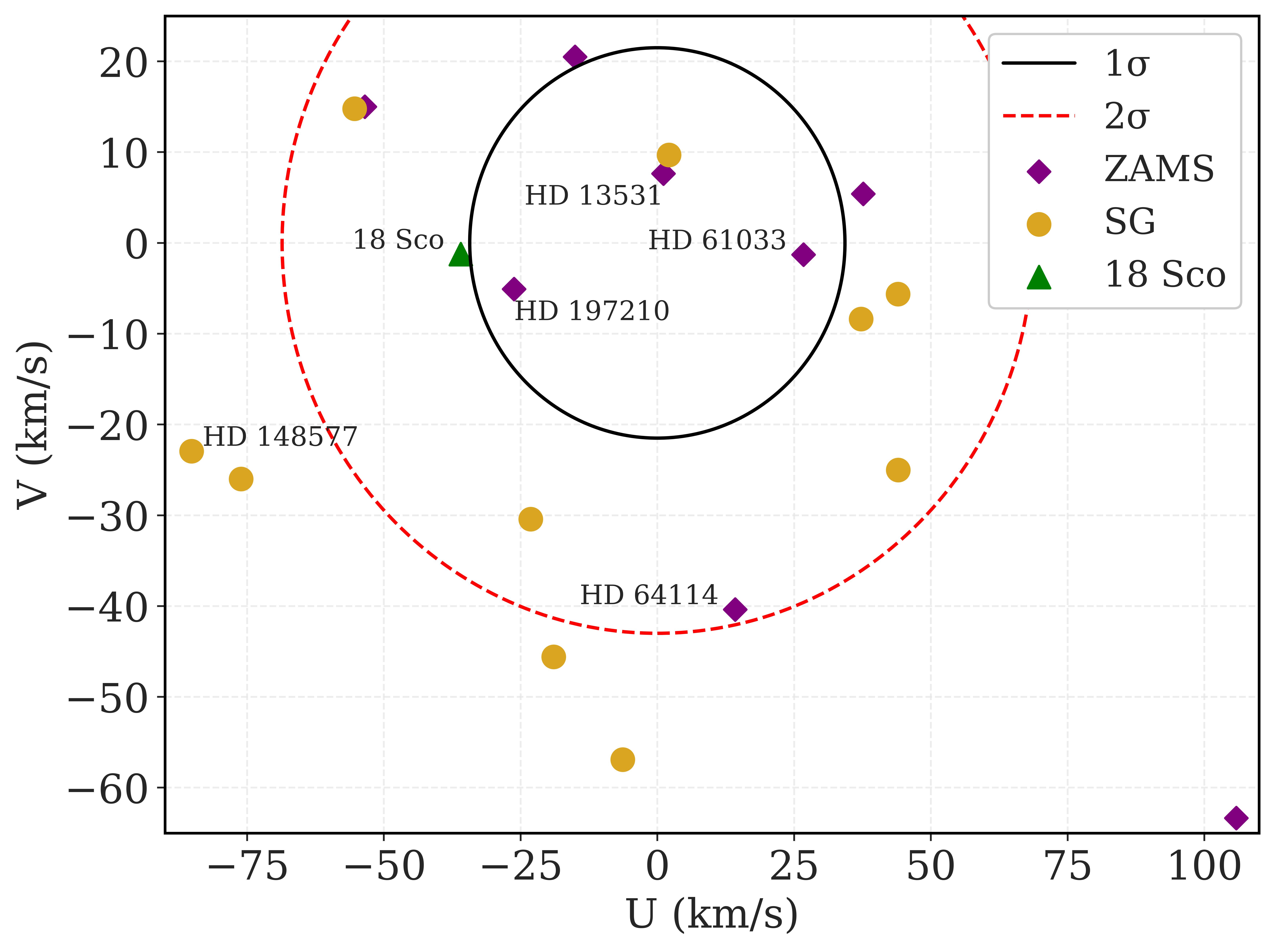} % Ajuste o caminho e o tamanho
    \includegraphics[width=0.45\textwidth]{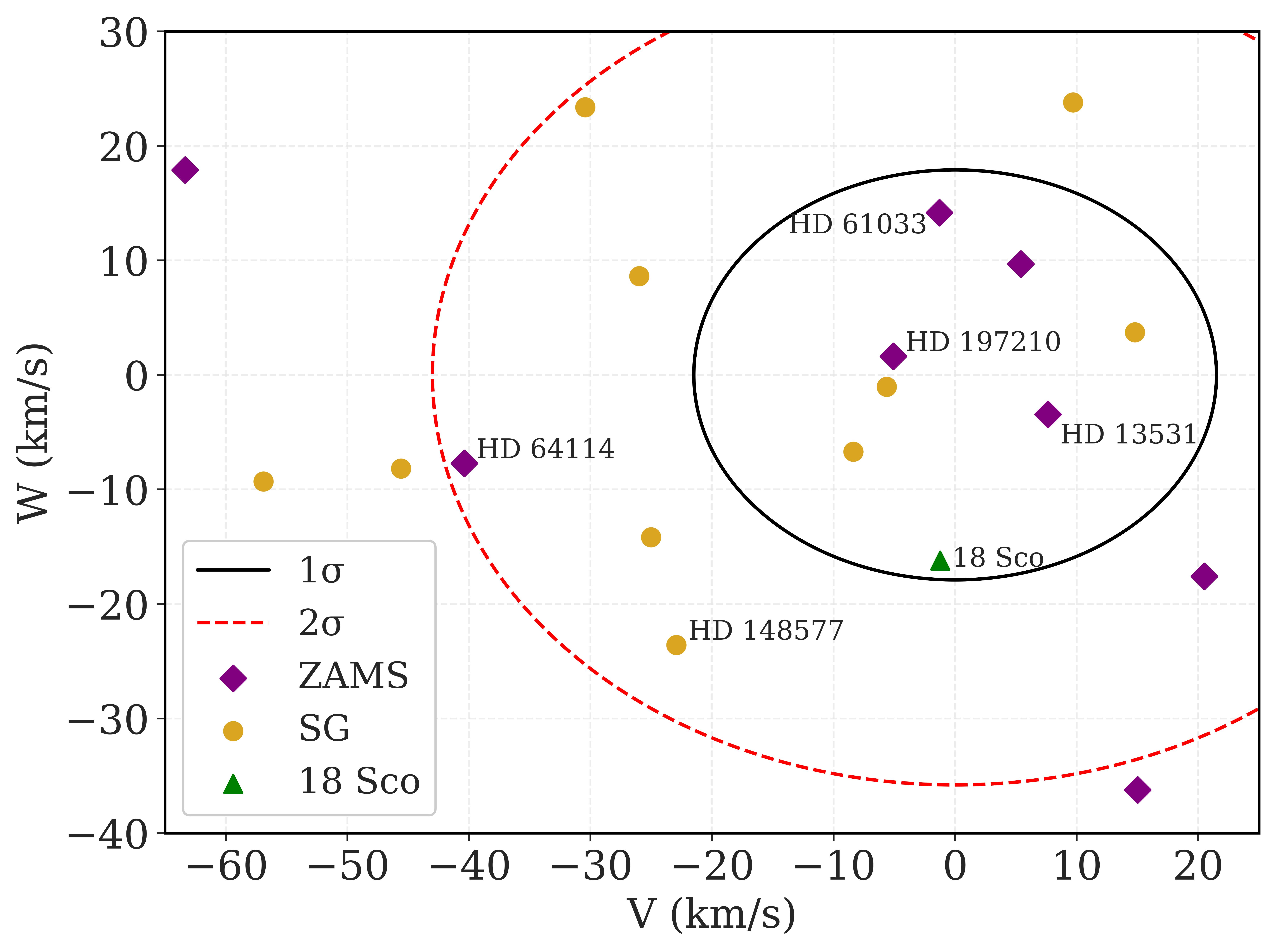} % Ajuste o caminho e o tamanho
    \caption{On the left panel, we present the diagram of the U x V components, while on the right panel, we show the diagram for the V x W components. The velocity dispersion ellipses for each component, with semi-major axes of 1$\sigma$ and 2$\sigma$, are represented by the solid black and dashed red lines, respectively. They were constructed based on the work of \citet{Turnbull_2003} and $\sigma$ values are: $\sigma_{U}$ = 34.3 $km s^{-1}$, $\sigma_{V}$ = 21.5 $km s^{-1}$ and $\sigma_{W}$ = 17.9 $km s^{-1}$. The purple diamonds represent the values of the ZAMS candidates, while the golden dots represent the SG candidates. Our control star 18 Sco is represented by the  green triangle. The labeled stars are discussed in further detail throughout the article.}
    \label{fig:velocity_components}
\end{figure*}

The \textit{UVW} velocity components of our candidates can be seen in Table \ref{table_param_kinematic}. It is important to note that the U component is positive towards the Galactic anticenter. The dispersion ellipses in the Figure \ref{fig:velocity_components} diagrams were created based on the work of \citet{Turnbull_2003}, in which velocity dispersions relative to the LSR were calculated for a significant sample of thin-disk stars with metallicity higher than –0.4. In general, we expect that ZAMS candidates should be more concentrated around the center of the dispersion ellipses, whereas SG candidates should exhibit greater dispersion in their components, which is what we observed in the diagrams of Figure \ref{fig:velocity_components}.

\begin{table}
    \centering
    \caption{Final kinematic parameters of the sample candidates. The columns in the table represent: star, stage to which it is a candidate and velocity components U, V and W, respectively.}
    \begin{tabular}{lccc}
        \hline
        Star         & \small{U} & \small{V} & \small{W}  \\
                     & $km s^{-1}$ & $km s^{-1}$ & $km s^{-1}$  \\
        \hline
        HD 13531 & 1.125   & 7.630  &-3.455\\
        HD 21411 & 37.634 & 5.411 & 9.686\\
        HD 25918 & -53.469 & 15.012 & -36.228\\
        HD 55720 & 105.849 & -63.345 & 17.890\\
        HD 61033 & 26.705  & -1.292  & 14.159 \\
        HD 64114 & 14.226  & -40.385  & -7.725 \\
        HD 182619 & -15.058 & 20.503 & -17.588\\
        HD 197210 & -26.206  & -5.089  & 1.619 \\
        HD 15942 & 37.259 & -8.382 & -6.709\\
        HD 19308 & 44.017 & -25.016 & -14.174\\
        HD 24040 & -18.968 & -45.585 & -8.172\\
        HD 69809 & 2.115 & 9.697 & 23.795\\
        HD 74698 & -23.148  & -30.433  & 23.370   \\
        HD 111398 & -55.336  & 14.797  & 3.713 \\
        HD 148577 & -85.143  & -22.943  & -23.580 \\
        HD 175425 &  -6.319 & -56.903 & -9.298\\
        HD 196050 &  -76.096 & -25.995 & 8.627\\
        HD 213575 & 43.987  & -5.636 & -1.035 \\
        18 Sco    & -35.926 & -1.238 & -16.214\\
        \hline
    \end{tabular}
\label{table_param_kinematic}
\end{table}

\section{Stellar activity and age diagnostics}
\label{sec:stellar_activity}

Stellar age is a fundamental parameter to understand the evolutionary stage of a star and it appears to be correlated with its level of stellar activiy. Furthermore, the latter can influence the atmospheres of exoplanets orbiting the star and their habitability, potentially affecting the retention of a sufficiently dense atmosphere due to atmospheric erosion processes (\citealt{powner2009synthesis,ribas2010evolution,krissansen2024erosion}), or even the development of conditions necessary for life as we know it. Therefore, an accurate determination of the stellar age is an essential parameter for a better understanding of the evolutionary scenario of the star and its system. The isochronal method (used in Section \ref{sec:evolparam}) has limitations, for example, in the ZAMS region the isochrones are close to each other resulting in a wider range of possible age solutions for a given set of input parameters. With respect to older stars, particularly the SG candidates, the isochrone method performs effectively due to the increased spacing between the isochrone curves, which facilitates an accurate age estimation. With that in mind, we decided to use other indicators, so we can estimate ages more accurately and better evaluate the evolutionary stage of each target star.

\subsection{\ion{Ca}{2} H \& K lines}
\label{sec:HK_lines}

The \ion{Ca}{2} H \& K lines can be good tracers for the age of a star because of their sensitivity to the star's chromospheric activity. The chromospheric flux contribution in these spectral lines can be measured using the S index (\citealt{Vaughan_1978}; \citealt{middelkoop1982magnetic}), which is the ratio of the flux in the line cores of the \ion{Ca}{2} H (3968.17{\AA}) \& K (3933.66{\AA}) lines and two nearby pseudo-continuum regions (3991.07 and 4011.07{\AA}). The S index can be used to obtain the $\text{log}{R'_{HK}}$ of each candidate, through proper flux calibration and photospheric correction, procedures that will be described in the next paragraphs. The $\text{log}{R'_{HK}}$ is the standard metric in the literature and we can apply activity-age relations to estimate the age of the candidates (e.g., \citealt{lorenzo2016age}). 

We used a code that compiles a database of S indexes from a variety of sources: \citet{henry1996survey}; \citet{gray2003contributions}; \citet{wright2004chromospheric}; \citet{gray2006contributions}; \citet{jenkins2008metallicities}; \citet{Isaacson_2010}; \citet{arriagada2011chromospheric} and \citet{jenkins2011chromospheric}. Nonetheless, we gave priority to \citet{Isaacson_2010} and \citet{wright2004chromospheric} surveys because they were constructed based on years of time series with the HIRES/Keck spectograph. For stars present in both of these catalogs, we adopted the average value. If they were present in neither of these surveys, the code returns the median value found on the other surveys. For the control star, HD 146233, we adopted the S index derived by \citet{lorenzo2018solar} through long-term monitoring of its chromospheric cycle.

\begin{figure}[h!]
    \centering
	\includegraphics[width=0.7\columnwidth]{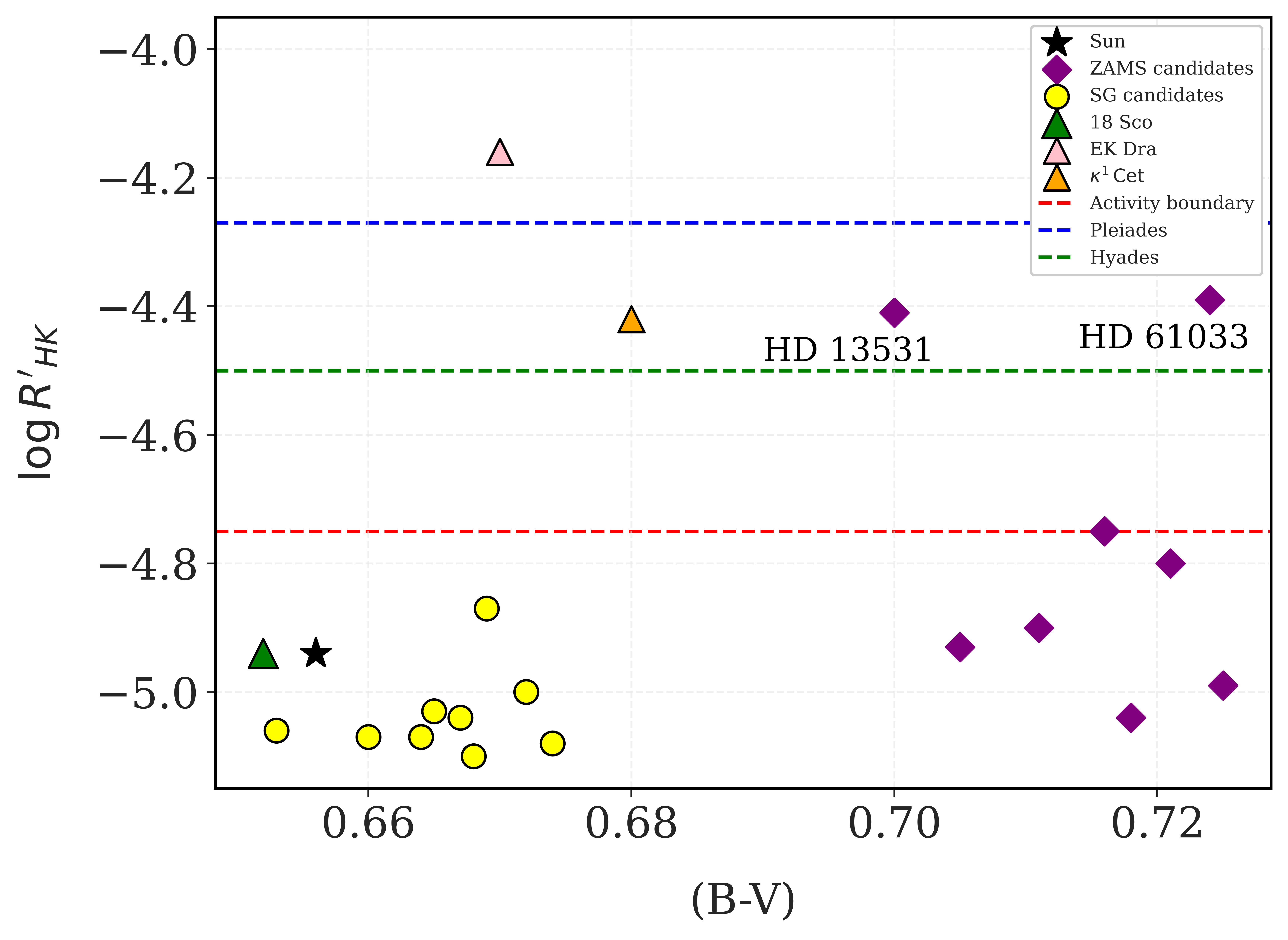}
    \caption{Chromospheric activity indicator $\text{log}{R'_{HK}}$ as a function of (B-V) for all stars in the sample. The purple diamonds represent the values of the ZAMS candidates, while the yellow points represent the SG candidates. Our control star 18 Sco is represented by the green triangle ($\text{log}{R'_{HK}} = -4.94$, as derived in Section \ref{sec:HK_lines}). The black star represents the values adopted for the Sun for comparison ($\text{log}{R'_{HK}} = -4.94$; \citealt{2018A&A...619A..73L}). The red dashed line represents the value of $\text{log}{R'_{HK}} = -4.75$ that was used as boundary to separate the active stars of the sample from the inactive ones (\citealt{henry1996survey}). The blue and green dashed lines show the average values for the Pleiades ($\text{log}{R'_{HK}} = -4.27$) and Hyades ($\text{log}{R'_{HK}}=-4.50$) active groups, respectively, derived by \citet{Mamajek_2008} for comparison. We also show the $\text{log}{R'_{HK}}$ values for the young sun analogs $\kappa^1 \text{ Cet}$ ($\text{log}{R'_{HK}} = -4.42$) and EK Dra ($\text{log}{R'_{HK}} = -4.16$) as the pink and orange triangles, respectively, from \citet{2022MNRAS.514.4300B} and \citet{Mamajek_2008}.}
    \label{fig:B_V_RHK}
\end{figure}

Finally, we converted the S indexes to $\text{log}{R'_{HK}}$ following the procedures described in \citet{rutten1984magnetic} and \citet{middelkoop1982magnetic}. In summary, the S index is affected by the spectral type of the star, as the measurement depends on the nearby continuum regions. Therefore, a correction must be applied to remove this effect (\citealt{middelkoop1982magnetic}). Additionally, it is necessary to eliminate the photospheric contribution (R$_{phot}$) present at the center of these lines within the spectral measurement window. The $\text{log}{R'_{HK}}$ index of each star is shown as a function of their (B-V) values in the Figure \ref{fig:B_V_RHK}. The (B-V) color values for each candidate were obtained from the Hipparcos catalog (\citealt{perryman1997hipparcos}). With the $\text{log}{R'_{HK}}$ estimated, we applied the age-mass-metallicity-activity relation from \citet{lorenzo2016age} to obtain the ages of our target stars, which is represented by the equation below:

\begin{equation}
\label{eq_activity-age}
%    \begin{split}
        \text{log} (\text{Age}_{HK}) =  -56.01 - 25.81 \cdot \text{log}(R'_{HK}) - 0.436\cdot\text{[Fe/H]} \\ - 1.26\cdot\text{log}(M/M_{\odot}) - 2.529\cdot\text{log}(R'_{HK})^2.
%    \end{split}
\end{equation}

The values adopted for the metallicity and mass were retrieved from Tables \ref{table_param_atm} and \ref{table_result_finais_complete}, respectively. The individual results for the chromospheric age can be seen in Table \ref{tab:converted_table}. As can be seen in Table \ref{tab:table_cromparam} and Figure \ref{fig:B_V_RHK}, the target stars HD 13531 and HD 61033 are very active and young, turning them into promising candidates to represent the young Sun, very similar to $\kappa^1 \text{ Cet}$ (\citealt{ribas2010evolution}), but less active and older when compared to the ZAMS analog EK Dra (\citealt{gudel1997x}). Furthermore, all other ZAMS candidates exhibit chromospheric activity levels below the threshold required to be classified as active, except for HD 21411, which lies precisely at the threshold ($\text{log}{R'_{HK}}$ = -4.75). Moreover, our best candidate for the SG stage (HD 148577) presents lower levels of chromospheric activity, which is consistent with older stars. Additionally, all the SG candidates are less active than the Sun, as expected, with the exception of the candidate HD 175425.

\begin{table}
	\centering
	\caption{Chromospheric age and input parameters used in the calculation. In order, the columns are: S index compiled from various literature sources (see Section \ref{sec:HK_lines}), the (B-V) color of each candidate.}
	\label{tab:table_cromparam}
	\begin{tabular}{lcccc} % four columns, alignment for each
		\hline
		Star    & $<S_{MW}>$ & $(B-V)$ & $\text{log}{R'_{HK}}$ \\
		\hline
		HD 13531 & 0.386 $\pm$ 0.005 & 0.700 & -4.41 \\
        HD 21411 & 0.224 $\pm$ 0.017 & 0.716 & -4.75 \\
        HD 25918 & 0.165 $\pm$ 0.005 & 0.725 & -4.99 \\
        HD 55720 & 0.175             & 0.705 & -4.93 \\
        HD 61033 & 0.409 $\pm$ 0.029 & 0.724 & -4.39 \\
        HD 64114 & 0.21 $\pm$ 0.01   & 0.721 & -4.80 \\
        HD 182619 & 0.156 $\pm$ 0.001 & 0.718 & -5.04 \\
        HD 197210 & 0.183 &  0.711  & -4.90  \\
        \hline
        HD 15942  &  -  & -  & -  \\
        HD 19308 & 0.160 $\pm$ 0.004 &  0.672 & -5.00 \\
        HD 24040 & 0.150 $\pm$ 0.003 &  0.653 & -5.06 \\
        HD 69809 & 0.148 $\pm$ 0.002 &  0.674 & -5.08 \\
        HD 74698 & 0.156 &  0.665  & -5.03  \\
        HD 111398 & 0.150 $\pm$ 0.002  &  0.660  & -5.07   \\
        HD 148577 & 0.152  &  0.664	 & -5.07  \\
        HD 175425 & 0.185  &  0.669  & -4.87 \\
        HD 196050 & 0.154 &  0.667 & -5.04 \\
        HD 213575 & 0.146  $\pm$ 0.001 & 0.668 & -5.10  \\
        \hline
        18 Sco  & 0.170 $\pm$ 0.004 & 0.652 & -4.94 \\
		\hline
	\end{tabular}
\end{table}

\subsection{The H$\alpha$ line}
\label{sec:Halpha}

Another interesting chromospheric indicator is the H$\alpha$ line ($\lambda$ = 6562.8 Å). Although chromospheric radiative losses are weaker in this line compared to the H \& K lines (\citealt{1991A&A...251..199P}), which can lead to higher relative uncertainties, H$\alpha$ fluxes have the advantage of being less affected by stellar magnetic cycles and transient phenomena such as flares and starspots (\citealt{2005A&A...431..329L}). These phenomena can cause variations in $\text{log}{R'_{HK}}$, even in very inactive stars (\citealt{2021A&A...646A..77G}). We compiled H$\alpha$ chromospheric fluxes (erg cm$^{-2}$ s$^{-1}$) for our sample stars following \citet{10.1093/mnras/stae1532}, who used atmospheric models and medium-resolution spectra. They also constructed an age-mass-metallicity-activity relation, expressed as:

\begin{equation}
%\begin{split}
    \text{log} (\text{Age}_{H\alpha}) = -107.306 + 42.976 \cdot \text{log} (F'_{H\alpha}) - 4.279 \cdot \text{log} (\text{M/M}_\odot)
     + 0.835 \cdot \text{[Fe/H]} - 3.931 \cdot \text{log} (F'_{H\alpha})^2 \ . 
%\end{split}
\label{eq:calib_age_ha}
\end{equation}

Their published chromospheric fluxes and ages are presented in Table \ref{tab:converted_table}, along with the corresponding uncertainties. The high H$\alpha$ chromospheric activity levels of HD 13531 and HD 61033, the only stars with F’$_{H\alpha}$ $>$ 10$^6$ erg cm$^{-2}$ s$^{-1}$, reinforce their youth. Among the subgiant candidates, all of them present F’$_{H\alpha}$ $<$ 10$^6$ erg cm$^{-2}$ s$^{-1}$, which is compatible with the hypothesis that they should have lower cromospheric activity levels.

\begin{table*}
    \centering
    \caption{In order, the columns are: the star identification, its proposed evolutionary stage, the H$\alpha$ flux F’$_{H\alpha}$ and its uncertainty in erg cm$^{-2}$ s$^{-1}$, the estimated age using H$\alpha$ activity relation (Age H$\alpha$), the age derived from the $\text{log}{R'_{HK}}$ index (Age HK), and the age from isochrone fitting (Age Iso) and their respective uncertainties. All age estimates are presented in gigayear units.}
    \begin{tabular}{lccccc}
        \hline
        ID & F’$_{H\alpha}$ & $\sigma$(F’$_{H\alpha})$ & Age$_{H\alpha}$ & Age$_{HK}$ & Age$_{Iso}$ \\
           & (erg cm$^{-2}$ s$^{-1})$ & (erg cm$^{-2}$ s$^{-1})$ & (Gyr) & (Gyr)    & (Gyr)\\
        \hline
        HD 13531 & 1.230E+06 & 1.588E+05 & $0.576^{+0.425}_{-0.245}$ & $0.4^{+0.2}_{-0.1}$ & $0.6^{+0.7}_{-0.4}$ \\
        HD 21411 & 7.563E+05 & 1.426E+05 & $4.440^{+3.276}_{-1.885}$ & $5.0^{+1.9}_{-1.4}$ & $4.5^{+0.9}_{-0.9}$ \\
        HD 25918 & 4.474E+05 & 1.271E+05 & $15.852^{+11.695}_{-6.730}$ & $7.4^{+2.9}_{-2.0}$ & $8.1^{+1.3}_{-1.4}$ \\
        HD 55720 & 5.227E+05 & 1.424E+05 & $10.778^{+7.952}_{-4.576}$ & $9.8^{+3.7}_{-2.7}$ & $12.5^{+1.5}_{-1.5}$ \\
        HD 61033 & 1.599E+06 & 1.635E+05 & $0.167^{+0.123}_{-0.071}$ & $0.4^{+0.1}_{-0.1}$ & $0.7^{+0.9}_{-0.5}$ \\
        HD 64114 & 6.586E+05 & 1.471E+05 & $7.098^{+5.237}_{-3.014}$ & $4.1^{+1.5}_{-1.1}$ & $2.1^{+0.8}_{-0.8}$ \\
        HD 182619 & 6.803E+05 & 1.304E+05 & $5.943^{+4.385}_{-2.523}$ & $7.9^{+2.9}_{-2.2}$ & $6.0^{+1.5}_{-1.6}$ \\
        HD 197210 & 9.040E+05 & 1.484E+05 & $2.223^{+1.640}_{-0.944}$ & $5.5^{+2.1}_{-1.5}$ & $2.4^{+0.8}_{-0.7}$ \\
        \hline
        HD 15942 & 5.170E+05 & 1.459E+05 & $9.243^{+6.819}_{-3.924}$ & - & $1.3^{+0.3}_{-0.4}$ \\
        HD 19308 & 5.599E+05 & 1.471E+05 & $7.558^{+5.576}_{-3.209}$ & $4.8^{+1.8}_{-1.3}$ & $3.9^{+0.6}_{-0.6}$ \\
        HD 24040 & 8.554E+05 & 1.569E+05 & $2.138^{+1.577}_{-0.908}$ & $4.6^{+1.7}_{-1.3}$ & $5.3^{+0.6}_{-0.4}$ \\
        HD 69809 & 6.988E+05 & 1.537E+05 & $4.228^{+3.119}_{-1.795}$ & $4.1^{+1.6}_{-1.1}$ & $3.8^{+0.3}_{-0.3}$ \\
        HD 74698 & 3.895E+05 & 1.366E+05 & $13.350^{+9.850}_{-5.668}$ & $5.2^{+2.0}_{-1.4}$ & $7.3^{+0.3}_{-0.3}$ \\
        HD 111398 & 5.565E+05 & 1.489E+05 & $7.813^{+5.764}_{-3.317}$ & $5.8^{+2.2}_{-1.6}$ & $7.9^{+0.4}_{-0.4}$ \\
        HD 148577 & 5.830E+05 & 1.549E+05 & $7.117^{+5.251}_{-3.022}$ & $7.0^{+2.6}_{-2.0}$ & $9.8^{+0.6}_{-0.6}$ \\
        HD 175425 & 6.393E+05 & 1.482E+05 & $5.337^{+3.938}_{-2.266}$ & $3.7^{+1.4}_{-1.0}$ & $4.8^{+0.9}_{-0.9}$ \\
        HD 196050 & 6.800E+05 & 1.609E+05 & $3.964^{+2.925}_{-1.683}$ & $4.0^{+1.6}_{-1.1}$ & $4.2^{+0.2}_{-0.4}$ \\
        HD 213575 & 5.406E+05 & 1.454E+05 & $6.457^{+4.764}_{-2.741}$ & $8.0^{+3.1}_{-2.2}$ & $11.1^{+0.3}_{-0.3}$ \\
        \hline
        HD 146233 & 6.159E+05 & 1.583E+05 & $5.702^{+4.207}_{-2.421}$ & $5.1^{+1.9}_{-1.4}$ & $2.3^{+0.6}_{-0.5}$ \\
        \hline
    \end{tabular}
    \label{tab:converted_table}
\end{table*}

\subsection{TESS light curves}
\label{sec:TESS_lightcurves}

\begin{figure}[h!]
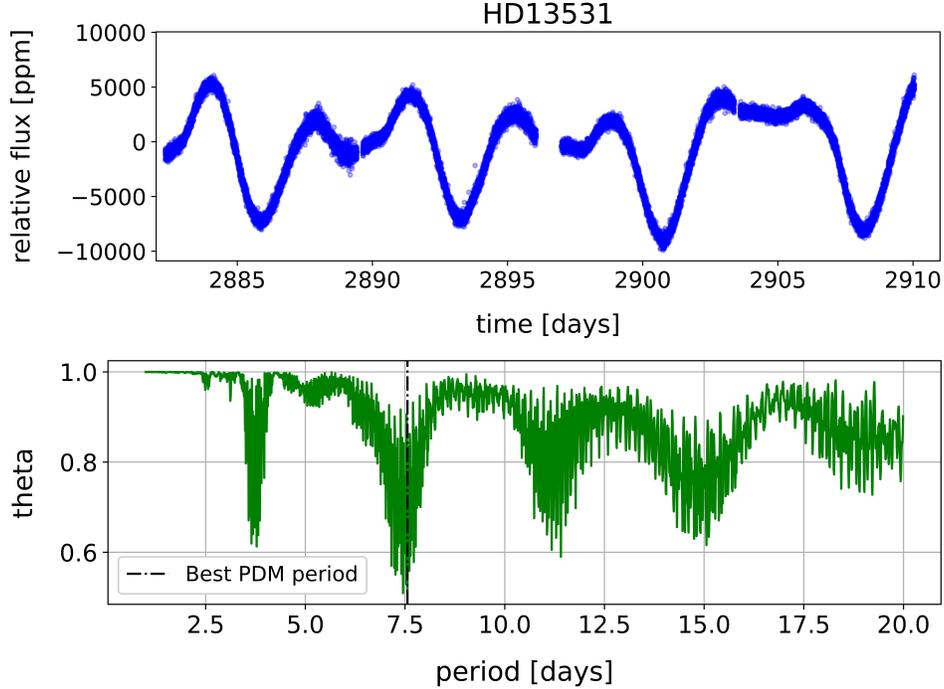

    \centering
	\includegraphics[width=0.7\columnwidth]{HD13531_curvadeluz.jpg}
    \includegraphics[width=0.7\columnwidth]{Ajuste_HD13531}
    \caption{In the top pannel, the light curve that was used to determine the rotation period of the star HD 13531. In the bottom panel, the minimum value for theta obtained from the PDM method (phase dispersion minimization; \citealt{jurkevich1971method}; \citealt{stellingwerf1978period}) is shown, corresponding to P$_{rot}$ = 7.6 days.}
    \label{fig:HD13531_lightcurve}
\end{figure}

The TESS (Transiting Exoplanet Survey Satellite/NASA\footnote{https://science.nasa.gov/mission/tess/}, \citealt{2010AAS...21545006R}) mission provides valuable photometric data for determining rotation periods of magnetically active stars. By delivering high-precision, short-cadence light curves, TESS enables the detection of rotational modulation caused by surface inhomogeneities, such as star spots. This technique is particularly effective for stars with strong magnetic activity, where the modulation amplitude is more pronounced, as illustrated by HD 13531 in Figure \ref{fig:HD13531_lightcurve} (upper panel). The continuous coverage of $\sim27$ days per sector allows the reliable determination of rotation periods, particularly for spot-dominated stars with moderate to rapid rotation ($P_{\mathrm{rot}} \lesssim 15$ days). According to standard gyrochronology relations (e.g., \citealt{2007ApJ...669.1167B}; \citealt{2008ApJ...687.1264M}), a single Sun-like star with $P_{\mathrm{rot}} \simeq 15$ days typically corresponds to an age between 1.5 and 2.5 Gyr. Therefore, photometric rotation periods could be only derived for the youngest and most active stars in our sample: HD 13531 and HD 61033.

We obtained their light curves using the \textit{lightkurve} package \citep{2018ascl.soft12013L}, and applied three independent techniques to estimate the rotation period: Lomb Scargle \citep{vanderplas2018understanding}, an auto-correlation function \citep{reinhold2020stellar} and the Phase dispersion minimization method (PDM; \citealt{jurkevich1971method}; \citealt{stellingwerf1978period}). In addition, we estimated their ages using the \textit{gyrointerp} package and their gyrochrones \citep{2023ApJ...947L...3B}. All the TESS data used in this paper can be found in \citealt{mast_tess_lightcurves_allsectors}: \dataset[https://doi.org/10.17909/t9-nmc8-f686]{https://doi.org/10.17909/t9-nmc8-f686} - Sectors: 7, 8, 9, 18, 34, 35, 36, 58, 85.

Figure \ref{fig:HD13531_lightcurve} shows, in the upper panel, the TESS light curve of the active star HD 13531. In the lower panel, we present the PDM analysis, where the green curve shows the theta values for different periods, which represent the variance within the phase bins. The period that minimizes theta corresponds to the true rotational period, indicated by the black line around 7.6 days. This result is consistent with the 7.5-day period reported in the GCVS catalog \citep{2017ARep...61...80S} and also with the stellar ages estimated from the Ca II H $\&$ K activity index, the H$\alpha$ method and the isochronal method (0.4, 0.576 and 0.6 Gyr, respectively). The ACF method yielded a rotation period of approximately 6.02 days, consistent wiht a young and active star. In addition, the Lomb-Scargle periodogram identified a strong signal at 3.74 days, which is likely a harmonic of the true $\approx$7-day rotational period. 

\begin{figure}[h!]
    \centering
	\includegraphics[width=0.7\columnwidth]{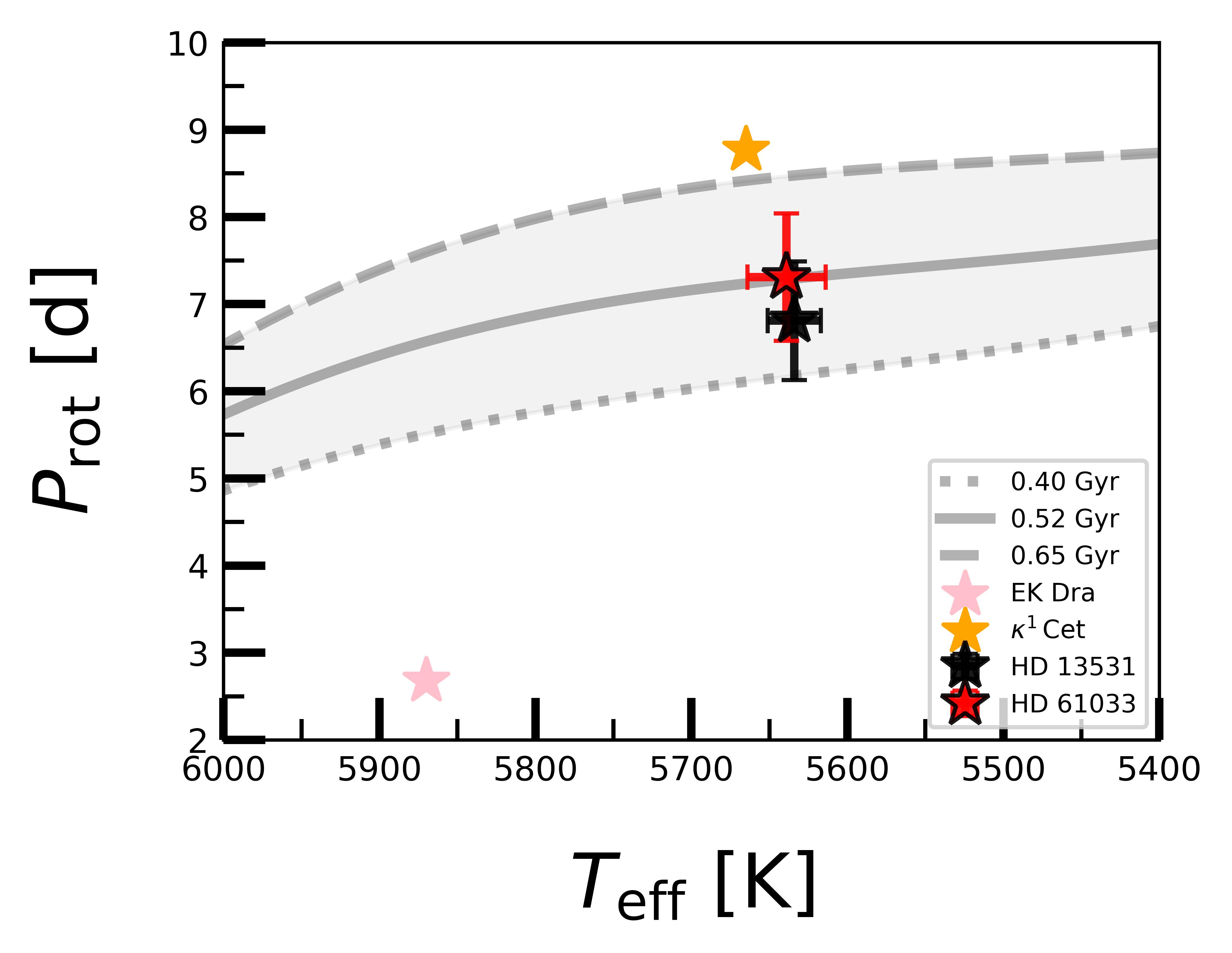}
    \caption{Rotation period ($P_{\mathrm{rot}}$) as a function of effective temperature ($\bm{T}_{eff}$) for the candidate stars. The grey lines represent gyrochrones for different ages: 0.40 Gyr (dotted), 0.52 Gyr (solid), and 0.65 Gyr (dashed). The black and red stars correspond to HD 13531 and HD 61033, respectively. Error bars reflect the uncertainties in both $\bm{T}_{eff}$ and $P_{\mathrm{rot}}$ measurements. The pink and orange stars represent EK Dra and $\kappa^1 \text{ Cet}$ respectively, with $\bm{T}_{eff}$ and $P_{\mathrm{rot}}$ adopted from \citealt{ribas2010evolution} and \citealt{gudel1997x}.}
    \label{fig:age_rot_fit}
\end{figure}

For the star HD 61033, another active candidate, the three methods used provide an average of P$_{rot}$ = 7.31$\pm$0.01. This value is compatible with a very active star, around 0.5 Gyr, a value compatible with its chromospheric and isochronal ages. In Figure \ref{fig:age_rot_fit} we present the final fit for the estimated ages using the mean value estimated by the three methods. For comparison, in Figure \ref{fig:age_rot_fit} we also present the rotational period of the young sun analogs $\kappa^1 \text{ Cet}$ (8.77 days - \citealt{2007ApJ...659.1611W};\citealt{ribas2010evolution}) and EK Dra (2.68 days - \citealt{gudel1997x}). The rotational period and age of our young sun candidates reinforce their candidacy, both of them are similar to $\kappa^1 \text{ Cet}$, however, older (and more evolved) than EK Dra.

\section{Best candidates}
\label{best_candidates}

Throughout Sections \ref{sec:stellar_parameters} and \ref{sec:stellar_activity}, we have performed a very detailed characterization of the candidates in our sample. Now, we will use the previous results to define which stars can be selected as interesting candidates to represent the Sun at different stages.

\begin{table}[h!]
    \centering
    \caption{Final evolutionary parameters for all the candidates. In order, the columns are: the star identification, the extinction in the V band $A_{V}$ in magnitudes, the stellar mass $M$ in solar units, the stellar radius $R$ in solar units, the stellar luminosity $L$ in solar units, the age derived from isochrone fitting ($\text{Age}_{iso}$), the age derived from the $\log R'_{HK}$ index ($\text{Age}_{HK}$), the age estimated from the H$\alpha$ activity relation ($\text{Age}_{H\alpha}$), and the age determined from stellar rotation ($\text{Age}_{Rot}$). All age estimates are given in gigayears units.}
    \begin{tabular}{lcccccccccc}
        \hline
       Star & A$_{V}$ & M & R & L & {$\text{Age}_{iso}$} & $\text{Age}_{HK}$ & $\text{Age}_{H\alpha}$ & $\text{Age}_{Rot}$\\
             & (mag) &  (M$_{\odot}$)& (R$_{\odot}$) & (L$_{\odot}$) & (Gyr)            & (Gyr)       & (Gyr)       & (Gyr)   \\
        \hline
        Sun ZAMS & - &1.00 & 0.90 & 0.70 & 0.0  & 0.0 & 0.0 & 0.0       \\
        HD 13531 & - &0.95$_{-0.01}^{+0.01}$ & 0.86$_{-0.01}^{+0.01}$ & 0.66$_{-0.01}^{+0.01}$  & 0.6$_{-0.4}^{+0.7}$  & \( 0.4^{+0.2}_{-0.1} \) & \( 0.576^{+0.425}_{-0.245} \) & 0.47 \\
        HD 21411 & -& 0.84 $_{-0.01}^{+0.01}$ & 0.81$_{-0.01}^{+0.01}$ & 0.52$_{-0.01}^{+0.01}$ & 4.5$_{-0.9}^{+0.9}$ & \( 5.0^{+1.9}_{-1.4} \) & \( 4.440^{+3.276}_{-1.885} \) & -            \\
        HD 25918 & -&  0.89$_{-0.01}^{+0.01}$ & 0.93$_{-0.02}^{+0.02}$ & 0.72$_{-0.02}^{+0.02}$ & 8.1$_{-1.4}^{+1.3}$ & \( 7.4^{+2.9}_{-2.0} \) & \( 15.852^{+11.695}_{-6.730} \) & -            \\
        HD 55720 & 0.019 & 0.80$_{-0.01}^{+0.01}$ & 0.89$_{-0.01}^{+0.02}$ & 0.66$_{-0.02}^{+0.02}$ & 12.5$_{-1.5}^{+1.5}$ & \( 9.8^{+3.7}_{-2.7} \) & \( 10.778^{+7.952}_{-4.576} \) & -             \\
        HD 61033 & 0.028 & 0.96$_{-0.01}^{+0.01}$ & 0.87$_{-0.01}^{+0.01}$ & 0.67$_{-0.02}^{+0.03}$ & 0.7$_{-0.5}^{+0.9}$  & \( 0.4^{+0.1}_{-0.1} \) & \( 0.167^{+0.123}_{-0.071} \) & 0.52  \\
        HD 64114 & - & 0.95$_{-0.01}^{+0.01}$ & 0.89$_{-0.01}^{+0.01}$ & 0.70$_{-0.03}^{+0.01}$  & 2.1$_{-0.8}^{+0.8}\)  & \( 4.1^{+1.5}_{-1.1} \) & \( 7.098^{+5.237}_{-3.014} \) & -   \\
        HD 182619 & 0.024 & 0.90$_{-0.01}^{+0.01}$ & 0.90$_{-0.01}^{+0.01}$ & 0.69$_{-0.02}^{+0.02}$  & 6.0$_{-1.6}^{+1.5}\)  & \( 7.9^{+2.9}_{-2.2} \) & \( 5.943^{+4.385}_{-2.523} \) & -              \\
        HD 197210 & - & 0.94$_{-0.01}^{+0.01}$ & 0.88$_{-0.01}^{+0.01}$ & 0.67$_{-0.01}^{+0.01}$ & 2.4$_{-0.7}^{+0.8}\)  & \( 5.5^{+2.1}_{-1.5} \) & \( 2.223^{+1.640}_{-0.944} \) & -               \\
        \hline
        Sun SG & -& 1.00 & 1.34 & 1.76  & 10.0  & 10.0 & 10.0 & 10.0\\
        HD 15942 & -& 1.22$_{-0.01}^{+0.01}$ & 1.19$_{-0.02}^{+0.02}$ & 1.57$_{-0.04}^{+0.05}$  & 1.3$_{-0.4}^{+0.3}\) &  -  & \( 9.243^{+6.819}_{-3.924} \) & -\\
        HD 19308 & - & 1.08$_{-0.01}^{+0.02}$ & 1.13$_{-0.02}^{+0.02}$ & 1.34$_{-0.04}^{+0.04}$  & 3.9$_{-0.6}^{+0.6}\) & \( 4.8^{+1.8}_{-1.3} \) & \( 7.558^{+5.576}_{-3.209} \) & -\\
        HD 24040 & - & 1.12$_{-0.02}^{+0.01}$ & 1.29$_{-0.02}^{+0.03}$ & 1.71$_{-0.05}^{+0.05}$  & 5.3$_{-0.4}^{+0.6}\) & \( 4.6^{+1.7}_{-1.3} \) & \( 2.138^{+1.577}_{-0.908} \) & -\\
        HD 69809 & - & 1.15$_{-0.01}^{+0.01}$ & 1.23$_{-0.01}^{+0.03}$ & 1.61$_{-0.04}^{+0.06}$ & 3.8$_{-0.3}^{+0.3}\) & \( 4.1^{+1.6}_{-1.1} \) & \( 4.228^{+3.119}_{-1.795} \) & -\\
        HD 74698 & 0.039 & 1.06$_{-0.01}^{+0.01}$ & 1.31$_{-0.02}^{+0.02}$ & 1.75$_{-0.05}^{+0.05}$  & 7.3$_{-0.3}^{+0.3}\) & \( 5.2^{+2.0}_{-1.4} \) & \( 13.350^{+9.850}_{-5.668} \) & -\\
        HD 111398 & - &  1.03$_{-0.01}^{+0.01}$ & 1.25$_{-0.02}^{+0.02}$ & 1.54$_{-0.05}^{+0.04}$  & 7.9$_{-0.4}^{+0.4}\) & \( 5.8^{+2.2}_{-1.6} \) & \( 7.813^{+5.764}_{-3.317} \) & -\\
        HD 148577 & -&  0.97$_{-0.01}^{+0.01}$ & 1.26$_{-0.02}^{+0.02}$ & 1.55$_{-0.04}^{+0.05}$  & 9.8$_{-0.6}^{+0.6}\) & \( 7.0^{+2.6}_{-2.0} \) & \( 7.117^{+5.251}_{-3.022} \) & -\\
        HD 175425 & -& 1.10$_{-0.02}^{+0.02}$ & 1.22$_{-0.03}^{+0.03}$ & 1.60$_{-0.05}^{+0.04}$  & 4.8$_{-0.9}^{+0.9}\) & \( 3.7^{+1.4}_{-1.0} \) & \( 5.337^{+3.938}_{-2.266} \) & -\\
        HD 196050 & 0.028 & 1.18$_{-0.01}^{+0.02}$ & 1.37$_{-0.02}^{+0.04}$ & 2.05$_{-0.04}^{+0.1}$  & 4.2$_{-0.4}^{+0.2}\) & \( 4.0^{+1.6}_{-1.1} \) & \( 3.964^{+2.925}_{-1.683} \) & -\\
        HD 213575 & - & 0.95$_{-0.01}^{+0.01}$ & 1.41$_{-0.02}^{+0.02}$ &1.86$_{-0.05}^{+0.05}$ & 11.1$_{-0.3}^{+0.3}\)& \( 8.0^{+3.1}_{-2.2} \) & \( 6.457^{+4.764}_{-2.741} \) & -\\
        \hline
        Sun Today & - & 1.00 & 1.00 & 1.00 & - & - & - & - \\
        HD 146233 (18 Sco) & - & 1.03$_{-0.01}^{+0.01}$ & 0.99$_{-0.01}^{+0.01}$ & 1.00$_{-0.04}^{+0.01}$ & 2.3$_{-0.6}^{+0.5}\) & \( 5.1^{+1.9}_{-1.4} \) & \( 5.702^{+4.207}_{-2.421} \) & - \\
        \hline
    \end{tabular}
\label{table_result_finais_complete}
\end{table}

\subsection{Young Sun analogs}
\label{sec:result_ZAMS}

The consistency between the parameters of the ZAMS candidates and the reference solar values at this stage can be seen in Table \ref{table_tick}. The candidates HD 13531 and HD 61033 are excellent candidates to represent the young Sun.

The star HD 13531 is one of the best candidates of our sample. Its atmospheric and evolutionary parameters are close to those adopted as reference values in this work within 2$\sigma$. It exhibits chemical abundances close to solar values when the associated uncertainties are taken into account. The average abundance difference, $\Delta$[X/H], was $-0.0493 \pm 0.035$. Its components of the Galactic space velocities are consistent with those of a young star. We observed a strong chromospheric emission at the centers of the Ca II H \& K lines, which is in agreement with the indexes found in the literature (\citealt{wright2004chromospheric}; \citealt{Isaacson_2010}), and also confirmed by our estimated $\text{log}{R'_{HK}}$ value. Furthermore, this candidate exhibits an F’$_{H\alpha}$ that is consistent with a young and active star. Finally, we estimate its age through 4 different methods, and we obtained a 1$\sigma$ agreement between them, which are compatible with the values expected for a young solar analog. Its derived rotational period (P$_{rot}$ = 6.81 days) is consistent with our age estimates for this candidate, reinforcing its youth.

The candidate HD 61033 is a very active star, with $\text{log}{R'_{HK}}$ = -4.39, and its estimated rotational period  (P$_{rot}$ = 7.31±0.01 days) is compatible with this value. Besides these factors, its atmospheric and evolutionary parameters are very close, within 2$\sigma$ for almost every parameter, to those adopted for the ZAMS Sun. The candidate HD 61033 presents chemical abundances that are consistent with solar values within the uncertainties. The average $\Delta$[X/H] for this candidate is $-0.001 \pm 0.063$. Its age indicators also point to a young star ($\approx$ 0.3-0.5Gyr), which turns it into one, if not the most, promising candidates in the sample. Its Age$_{H\alpha}$ estimate indicates an even younger candidate (0.167 Gyr), slightly older than the Pleiades Cluster (\citealt{prvsa2016nominal}). This star would thus be one of our best candidates. However, during the course of this work, the candidate HD 61033 was identified as a spectroscopic binary (\citealt{coralie}). Its companion is a $0.32M_\odot$ star in an orbit with a semi-major axis of $0.91 \mathrm{AU}$.

We note that the candidate HD 13531 has ages of $\approx$ 0.3-0.5 Gyr, which places it as interesting target to study the conditions present on the planetary system at the epoch when life arose on Earth (Archaean Eon). These conditions are critical to the development of Earth's primitive atmosphere and the development of life itself. In addition, the candidate HD 197210 is quite interesting for the study of Sun-like stars, however, a little more evolved past the ZAMS, close to the age of 2.0 Gyr, when oxygen became significantly abundant in Earth's atmosphere.

\subsection{Subgiant stage}
\label{sec:result_SG}

The star HD 148577 has a marginally lower mass and higher metallicity than the Sun. Its effective temperature and surface gravity are inside the 1$\sigma$ interval around the theoretical point defined for the SG Sun. Its abundances are consistent with solar values within the uncertainties, with an average $\Delta$[X/H] of $0.0791 \pm 0.0835$. Its isochronal age is consistent with expectations for a star at this stage, while its chromospheric age is also relatively high, but lower than the isochronal estimate. This is confirmed by the nearly identical age estimates from (Age${H\alpha}$) and (Age${HK}$), indicating that this candidate is indeed fairly inactive and evolved, as expected, however, at a stage prior to the turnoff. We know that determining a precise chromospheric age for evolved stars is difficult, mainly due to the abrupt drop in chromospheric activity as stars evolve. \citet{lorenzo2016age} showed that this type of analysis can be done with stars up to at least $\sim$ 6 Gyr with good precision, which encompass almost all of our sample. In other words, we consider the isochronal age as the most reliable at this stage. We, therefore, consider this star to be an excellent candidate to represent the Sun at the SG stage.

\section{Conclusions}
\label{sec:conclusions}

In this work, we analyzed 18 candidates for representing the Sun at two stages of its evolution: ZAMS and subgiant (SG). We utilized high-resolution spectra (R $>$ 35000) with high signal-to-noise ratio (SNR $>$ 100) to derive the atmospheric parameters of all candidates through the classical spectroscopic method, which uses equivalent widths of Fe I and Fe II lines and is based on the excitation and ionization equilibria. We calculated the evolutionary parameters of these stars by comparing stellar parameters with grid of isochrones. Additionally, we determined the kinematic parameters of each candidate and estimated their ages using three additional indicators: chromospheric emission in the H \& K lines of Ca II, chromospheric emission in the ${H\alpha}$ line and their rotation periods derived from their TESS light curves. Our methods allowed us perform a very detailed characterization of each of the sample candidates and select those that could represent the Sun at different stages of its life.

In particular, we should highlight the following candidates: HD 13531 (young Sun analog), HD 61033 (young Sun analog) and HD 148577 (SG). These stars are the most promising in the sample, with the first two stars being very active and practically indistinguishable from the Sun around its first 0.5 Gyr. Besides, these two candidates present a good agreement between the 4 different age indicators used in this work (isochronal, two chromospheric and rotational ages). With ages in the interval $\approx$ 0.3 - 0.5Gyr, the young Sun candidate HD 13531 is very interesting, from an astrobiological perspective, because it could represent the Sun when life appeared on Earth (Archaean Eon). Therefore, this candidate is useful for a study of the habitability conditions in this type of stellar system, as well as priority target to search for exoplanets in future missions, when we will be able to detect Earth-sized planets in their habitable zones, for example, with the PLATO mission(\citealt{2025ExA....59...26R}).

The star HD 148577, on the other hand, is an excellent candidate to represent the Sun at the end of the Main Sequence. Its isochronal age suggests that this star is close to the values expected for the turn-off point, in addition to having already undergone a considerable increase in its radius and luminosity values. From an astrobiological perspective, this candidate presents a unique opportunity for studying and monitoring environmental conditions as a possible host of old, evolved exoplanets.

Our pilot study demonstrated that the selection method used in this work is capable of identifying Sun-like stars at different evolutionary stages, broadening the understanding gained through the solar twins approach. Currently, we are expanding our sample through a new selection of candidates leveraging Gaia photometric data \citep{gaia2018gaia} as well as the more updated PARSEC evolutionary tracks (\citealt{bressan2012parsec}). These tools will allow us to perform a more comprehensive study and better understand the solar evolution and habitability in nearby planetary systems.

%% Please use the acknowledgment and contribution environments. This will 
%% be anonomyized when the "anonymous" style option is used. 
\begin{acknowledgments}
E.-O.C.S. would also like to express my gratitude to Conselho Nacional de Desenvolvimento Científico e Tecnológico (CNPq) for financially supporting the project. Additionally, E.-O.C.S. would like to thank Fundação Coordenação de Aperfeiçoamento de Pessoal de Nível Superior (CAPES) for funding during my PhD (under grant number 88887.927885/2023-00), which enabled the development of this article. L.G. would like to thank the financial support from Fundação de Amparo à Pesquisa do Estado do Rio de Janeiro (FAPERJ) through the ARC grant number E-26/211.386/201. G.F.P.M. acknowledges financial support from CNPq/Brazil under grant 474972/2009-7. P.V.S.S. acknowledges CAPES/Brazil PhD scholarship under grant 88887.821758/2023-00. E.-O.C.S. and L.G. would like to thank Ignasi Ribas for his helpful suggestions regarding the final draft of the manuscript. We thank the staff of OPD/LNA for their considerable support during the many observation runs carried out during this project. The software IRAF (\citealt{1986SPIE..627..733T,1993ASPC...52..173T}) was used throughout this work - IRAF is listed in the Astronomical Source Code Library as ascl:9911.002; The DOI is 10.5281/zenodo.5816743.
\end{acknowledgments}

\begin{contribution}
%%This section gives authors the space to recognize author contributions. The text inside this environment is NOT counted towards the total word quanta. At a minimum, manuscripts are expected to include this text:

%All authors contributed equally to the Terra Mater collaboration.

%% But authors are expected to provide more specific details, e.g. 
%%
%%SC was responsible for writing and submitting the manuscript.
%%WWM came up with the initial research concept and edited the manuscript.
%%OTS obtained the funding and edited the manuscript.
%%EBF provided the formal analysis and validation. He also edited the manuscript.
%%GEH Supervised the undergraduates, wrote the software and administers the project github and Zenodo repositories.
%%
%% Authors can use the Contributor Role Taxonomy (CRediT) at
%% for ideas on how write a good statement tailored to their needs.

\end{contribution}

\facilities{MUSICOS(OPD/LNA)}

\software{\textit{NumPy} (\citealt{harris2020array}),
        \textit{Matplotlib} (\citealt{Hunter:2007})
          }


\appendix

\begin{table*}
    \centering
    \caption{Consistency between the stellar parameters and the reference solar values at each stage. The columns in the table represent: star, the evolutionary stage for which the candidate was selected, effective temperature, logarithm of the surface gravity, metallicity, mass, radius, luminosity, chromospheric activity index ($\text{log} R'{_{HK}}$), isochrone age, ages derived from chromospheric activity indices ($\text{log}R'_{HK}$ and H$\alpha$), rotational period, and kinematic properties. Ticks indicate that the parameters are consistent with the expected values within 2$\sigma$. We use the "$\approx$" symbol for values that were beyond but still close (less or equal to 0.02) to the 2$\sigma$ limits. For the effective temperatures ($\bm{T}_{eff}$), we adopted a threshold of up to 10K relative to the 2$\sigma$ limits. For ages, due to the large uncertainties in some measurements, the tick/X classification was based only on the central value and an approximate 1 Gyr interval.}
    \begin{tabular}{ccccccccccccccc}
    \hline
        Star         & Candidate & $\bm{T}_{eff}$   & $\text{log}$ $g$   & $[Fe/H]$ & M & R & L &  $\text{log}$R'$_{HK}$  & {$\text{Age}_{iso}$} & $\text{Age}_{HK}$ & $\text{Age}_{H\alpha}$ & P$_{rot}$ & Kinematics\\
        \hline
        Sun ZAMS & & 5586 & 4.53 & 0.00 & 1.00 & 0.90 & 0.70 & $>$-4.75 & 0.0 & 0.0 & 0.0 & - & - \\
        HD 13531 & ZAMS & $\times$ & $\checkmark$ & $\checkmark$ & $\times$ & $\approx$ & $\approx$ & $\checkmark$ & $\checkmark$ & $\checkmark$ & $\checkmark$ & $\checkmark$ & $\checkmark$\\
        HD 21411 & ZAMS & $\times$ & $\checkmark$ & $\times$ & $\times$ & $\times$ & $\times$ &  $\checkmark$ & $\times$ & $\times$ & $\times$ & - & $\checkmark$\\
        HD 25918 & ZAMS & $\approx$ & $\checkmark$ & $\approx$ & $\times$ & $\checkmark$ & $\checkmark$ & $\times$ & $\times$ & $\times$ & $\times$ & - & $\times$\\
        HD 55720 & ZAMS & $\times$ & $\checkmark$ & $\times$ & $\times$ & $\checkmark$ & $\checkmark$  & $\times$ & $\times$ & $\times$ & $\times$ & - & $\times$ \\
        HD 61033 & ZAMS & $\approx$ & $\checkmark$ & $\checkmark$ & $\approx$ & $\approx$ & $\checkmark$ & $\checkmark$ & $\checkmark$ & $\checkmark$ & $\checkmark$ & $\checkmark$ & $\checkmark$\\
        HD 64114 & ZAMS & $\checkmark$ & $\checkmark$ & $\checkmark$ & $\times$ & $\checkmark$ & $\checkmark$ & $\times$ & $\times$ & $\times$ & $\times$ & - & $\times$\\ 
        HD 182619 & ZAMS & $\checkmark$ & $\checkmark$ & $\times$ & $\times$ & $\checkmark$ & $\checkmark$  & $\times$ & $\times$ & $\times$ & $\times$ & - & $\checkmark$\\
        HD 197210 & ZAMS & $\checkmark$ & $\checkmark$ & $\checkmark$ & $\times$ & $\checkmark$ & $\approx$ & $\times$ & $\times$ & $\times$ & $\times$ & - & $\checkmark$\\
        \hline
        Sun SG & & 5743 & 4.18 & 0.00 & 1.00 & 1.34 & 1.76 & $<$-4.75 & 10.0 & 10.0 & 10.0 & - & -\\
        HD 15942 & SG & $\times$ & $\times$ & $\times$ & $\times$ & $\times$ & $\times$ & $\checkmark$ & $\times$ & - & $\checkmark$ & - & $\checkmark$\\
        HD 19308 & SG & $\times$ & $\times$ & $\times$ & $\times$ & $\times$ & $\times$ & $\checkmark$ & $\times$ & $\times$ & $\times$ & - & $\checkmark$\\
        HD 24040 & SG & $\times$ & $\approx$ & $\times$& $\times$ & $\checkmark$ & $\checkmark$ & $\checkmark$ & $\times$ & $\times$ & $\times$ & - & $\checkmark$ \\
        HD 69809 & SG & $\times$ & $\times$ & $\times$ & $\times$ & $\times$ & $\checkmark$ & $\checkmark$ & $\times$ & $\times$ & $\times$ & - & $\checkmark$\\
        HD 74698 & SG & $\times$ & $\checkmark$ & $\times$ & $\times$ & $\checkmark$ & $\checkmark$ & $\checkmark$ & $\times$ & $\times$ & $\times$ & - & $\checkmark$\\
        HD 111398& SG & $\checkmark$ & $\times$ & $\times$ & $\approx$ & $\times$ & $\times$ & $\checkmark$ & $\times$ & $\times$ & $\times$ & - & $\checkmark$\\
        HD 148577& SG & $\checkmark$ & $\checkmark$ & $\checkmark$ & $\approx$ & $\times$ & $\times$  & $\checkmark$ & $\checkmark$ & $\times$ & $\times$ & - & $\checkmark$\\
        HD 175425& SG & $\times$ & $\checkmark$ & $\times$ & $\times$ & $\times$ & $\times$ & $\checkmark$ & $\times$ & $\times$ & $\times$ & - & $\checkmark$\\
        HD 196050& SG & $\times$ & $\checkmark$ & $\times$ & $\times$ & $\checkmark$ & $\times$ & $\checkmark$ & $\times$ & $\times$ & $\times$ & - & $\checkmark$\\
        HD 213575& SG & $\times$ & $\checkmark$ & $\times$ & $\times$ & $\times$ & $\checkmark$ & $\checkmark$ & $\checkmark$ & $\times$ & $\times$ & - & $\checkmark$\\
        \hline
    \end{tabular}
\label{table_tick}
\end{table*}

%\being{rotatetable}

%\end{rotatetable}

%% For this sample we use BibTeX plus aasjournalv7.bst to generate the

\bibliography{sample701}{}
\bibliographystyle{aasjournalv7}



\end{document}